\documentclass[%
 aps,
 sd,%
 amsmath,amssymb,
reprint,%
]{revtex4-1}

\usepackage{color}
\usepackage{graphicx}	
\usepackage{dcolumn}
\usepackage{bm}
\usepackage{dblfloatfix}

\newcommand\simlt{\lower.5ex\hbox{$\; \buildrel < \over \sim \;$}}
\newcommand\simgt{\lower.5ex\hbox{$\; \buildrel > \over \sim \;$}}

\begin{document}
\title{Existence of  steady gap solutions in rotating black hole magnetospheres}
\author{Amir Levinson \& Noam Segev}
\affiliation{Raymond and Beverly Sackler School of Physics \& Astronomy, Tel Aviv University, Tel Aviv 69978, Israel}

\begin{abstract}
Under conditions prevailing in certain classes of compact astrophysical systems, the active magnetosphere
of a rotating black hole becomes charge-starved, giving rise to formation of a spark gap in which plasma
is continuously produced.   The plasma production process 
is accompanied by curvature and inverse Compton emission
of gamma rays in the GeV-TeV band, that may be detectable by current and future 
experiments.    The properties of the gap emission have been studied recently using a fully general relativistic model 
of a local steady gap.  However, this model requires artificial adjustment of the electric current which is determined,
in reality, by the global properties of the magnetosphere. 
In this paper we map the parameter regime in which 
steady gap solutions exist, using a steady-state gap model in Kerr geometry, and show that such solutions are allowed only
under restrictive conditions that may not apply to most astrophysical systems.   We further argue that even the allowed 
solutions are inconsistent with the global magnetospheric structure.  We conclude that magnetospheric gaps are inherently intermittent,
and point out that this may drastically change their emission properties. 
\end{abstract}


\maketitle

\section{Introduction}
A question of considerable interest in the theory of Poynting-flux outflows from black holes (BHs) \cite{BZ77,BK08} is the nature 
of the plasma source in the magnetosphere. In difference from pulsars, in which free charges can be 
supplied to the magnetosphere by the rigid star 
along magnetic field  lines that are anchored to its surface,
in Kerr BHs there is no such an inherent  plasma source.   
As discussed in some greater detail in the next section, plasma in the region enclosed between the 
inner and outer Alfven surfaces must be continuously replenished by either some external agent
or via pair cascades in a spark gap. 

It has been argued that under conditions likely to prevail in many BH systems, both supermassive and stellar,
formation of a spark gap is inevitable  \cite{le00,LR11,HP16,Hir16}.   It has been further pointed out that the gap activity may be imprinted 
in the high-energy emission observed in these sources \cite{LR11,HP16,Hir16,Hir17,Lin17,NA07,Rie11,BT15}.  The variable TeV emission detected in M87 \cite{ahr03,acc09}, a galaxy that 
harbours one of largest BHs in the universe, as well as in the the radio galaxy IC 310 \cite{alk14}, has been regarded as being a plausible example of the signature 
of magnetospheric plasma production on horizon scales \cite{LR11,HP16,NA07}.   

In essence, the gap is an inherent part of the global magnetospheric structure.  Hence, a self-consistent analysis of magnetic outflows requires
a proper account of the coupling between the gap and the force-free regions of the outflow.   This can only be achieved,  at least in principle, using
global plasma simulations.  While global PIC simulations have been performed recently for pulsars \cite{CB14,CPPS15,PSC15},  they are expected to be far more 
involved in the case of black holes, since (i) a fully general relativistic scheme must be implemented, and  (ii) unlike in 
pulsars, the origin of the magnetic field threading the BH is poorly understood, which reflects on the choice of boundary conditions.   
To avoid such complications, and still get some insight into the physics underlying plasma production in the gap, local gap 
solutions can be sought, in which the global magnetospheric structure is assumed to be unaffected by the gap activity, while the 
magnetospheric current is treated as a free input  parameter of the gap model. 

In a recent series of papers \cite{HP16,Hir16,Hir17}, a fully general relativistic model of a steady gap has been developed and exploited to 
study the properties of magnetospheric emission.   In this model the magnetospheric current was not treated as a free parameter, but 
rather adjusted, for any given choice of the remaining parameters, to keep the multiplicity at the value required by the closure condition (c.f., Eq. (26) in Ref~\cite{Hir16}). 
The question then arises as to how restrictive are the conditions under which steady state solutions exist.  This issue is of importance, as
it might have drastic implications for the gap emission.
The point is that in steady gaps that  encompass the null surface the maximum power that can be released scales as $h^4$ with the gap width $h$ \cite{HP16}.
Since the pair multiplicity in a steady gap cannot exceed unity, this implies that  
the gamma-ray luminosity emitted from a steady gap decreases rapidly as the intensity of the external radiation source, that provides the pair production 
opacity, increases.  Such restrictions do not apply to intermittent gaps that can support a large magnetospheric current even when exposed to an intense
radiation field.  What are the limits on the output power of intermittent gaps is unclear at present.   Future plasma simulations might be able to resolve this question.

In this paper we map the parameter regime in which local steady gap solutions exist, using a 1D model of a local magnetospheric gap in Kerr geometry.   We find that such solutions require highly restrictive conditions, that may not apply to most astrophysical systems.  Moreover, we argue that even the local steady solutions that are allowed in this model are inconsistent with the global magnetospheric structure.    This implies that the plasma production region is dynamic, which may have far reaching consequences for the gap emission.  In Sec. \ref{sec:vac_break} we review the conditions under which gap formation is expected.  In Sec. \ref{sec:analysis} we present the model, and in Sec. \ref{sec:results} discuss
the results.  In sec. \ref{sec:remark} we briefly remark on the connection between the local model and the global structure.  We conclude in Sec. \ref{sec:conc}.

\section{Conditions for vacuum breakdown}
\label{sec:vac_break}
An inherent feature of MHD outflows driven 
by a Kerr BH is the presence of a stagnation surface located
between the inner and outer light cylinders (e.g., Refs~\cite{GL13,GL14,BT15,HP16}).
The  reason is that the strong  gravitational  field  of  the  black  hole
imposes an inward motion of plasma very near the horizon,
regardless of the direction of the energy flux, whereas the plasma above the outer light 
cylinder must be flowing outwards.  Consequently, the plasma in the causal magnetospheric region 
must be continuously replenished.

The  injection  of  charges  into  the  magnetosphere  may  be
associated  with  the  accretion  process.  Direct  feeding  seems
unlikely, as charged particles would have to cross magnetic field
lines on a timescale shorter than the accretion time in order to
reach the polar outflow.  Magnetic field irregularities, either inherent 
or forming by some macroscopic instabilities, can give rise to occasional
loading of the magnetosphere.  However, the timescale of such episodes may 
be considerably longer than the escape time of plasma in the inner magnetosphere 
(around the stagnation surface), so that some additional injection process may 
be required to maintain the local charge density above the Goldreich-Julian (GJ) value
everywhere in the magnetosphere.
In AGNs and microquasars this may be accomplished through annihilation of MeV photons
emanating from the hot gas accreted into the black hole.  We denote the luminosity of
this radiation source, henceforth measured in Eddington units, by $l_\gamma=L_\gamma/L_{Edd}$, 
and its size, given in units of $r_g$, by $\tilde{R}_\gamma=R_\gamma/r_g$.
For sufficiently high annihilation rate the resultant  charge  density  can  exceed  the  GJ  value,
keeping  the  magnetosphere  force-free.  At  lower  annihilation
rates the magnetosphere will be starved and a gap should form. 

The density of injected pairs can be estimated by equating the pair production rate with the escape rate.
It is given roughly by $n_\pm\simeq \sigma_{\gamma\gamma}n^2_\gamma r_g/3$ \cite{LR11}, where 
$n_\gamma\simeq 10^{22}\, m^{-1}\tilde{R}_\gamma^{-2} l_\gamma$  cm$^{-3}$ is the density of MeV photons, and $m=M_{BH}/M_\odot$
is the black hole mass in solar mass units.  Complete screening requires $n_\pm > n_{GJ}$, 
here $n_{GJ}=\Omega B/(2\pi e c)=2\times10^{11} B_8(\Omega/\omega_H)m^{-1}$ cm$^{-3}$ 
denotes the GJ density, $\Omega$ is the angular velocity of magnetic surfaces, $\omega_H\simeq c/2r_g$ is the 
angular velocity of the black hole, $B=10^8 B_8$ Gauss is the strength of the magnetic field near the horizon, 
and $e>0$ is the magnitude of the electron charge. The later condition can be expressed as:  
\begin{equation}
l_\gamma > 10^{-3} B_8^{1/2}(\Omega/\omega_H)^{1/2} (\tilde{R}_\gamma/30)^2.
\end{equation}
For smaller values of $l_\gamma$ the magnetosphere becomes charge starved and a gap forms.

The strength of the magnetic field near horizon can be estimated by assuming that it is in rough equipartition with the ram pressure in the disk.
This yields 
\begin{equation}
B\simeq 10^9 \dot{m}^{1/2} m^{-1/2}\quad {\rm G},
\end{equation}
where $\dot{m}=\eta\dot{M}c^2/ \dot{L}_{Edd}$, with $\eta\simeq0.1$ being the radiative efficiency, is the acrretion rate in Eddington units.

In  the RIAF regime the accretion flow is hot and the gamma ray luminosity can be estimated from an ADAF model, e.g., Ref~ \cite{NY95}, 
up to some uncertainty in the electron temperature.   Adopting such a model yields a condition for the appearance of a gap:
$\dot{m} <4\times10^{-3} m^{-1/7}$ \cite{LR11,Hir16}.  At higher accretion rates  the accretion disk spectrum cannot extend to high energies,
as it is too cold.  However, gamma-rays may originate from a tenuous corona, if present as widely believed, although no reliable 
constraints on the spectrum and luminosity of this coronal component have been imposed thus far.  In principle, it could be that in sources 
that accrete at relatively high rates the magnetic field is much higher than in RIAF sources, while the gamma ray luminosity is suppressed. 
If indeed true it could mean that gap emission in such objects may be more intense than in RIAF sources.

\section{\label{sec:analysis}A stationary gap model}

We construct a model describing a 1D, general relativistic stationary gap, that treats the electron-positron plasma as a two-beam fluid. 
The global magnetic field geometry adopted below is a split monopole geometry.
The gap extends along a poloidal magnetic surface, characterized by an inclination angle $\theta$. 
Gamma rays are produced by accelerating pairs via curvature emission and inverse Compton (IC) scattering, and in turn generate fresh pairs
through their interaction with an ambient radiation field, given as input.    It should be emphasized that these local steady gap solutions
are applicable only in the region where ideal MDH breaks down.  In the global picture additional forces are acting on the particles that
are ignored here, which will determine the conditions outside the gap.  The details are outlined in the following:
\subsection{Background geometry}
The background spacetime is described by the Kerr metric, here 
given in Boyer-Lindquist cordinates with the following notation:
\begin{equation}
ds^2= -\alpha^2dt^2 + g_{\varphi\varphi}(d\varphi-\omega dt)^2 + g_{rr}dr^2 + 
g_{\theta\theta}d\theta^2,
\end{equation}
where 
\begin{eqnarray}
 \alpha^2&=&\frac{\Sigma\Delta}{A};\quad \omega=\frac{2ar_g r}{A};\quad g_{rr}=\frac{\Sigma}{\Delta};\\
&g_{\theta\theta}&=\Sigma; \quad g_{\varphi\varphi}=\frac{A}{\Sigma}\sin^2\theta, \nonumber
\end{eqnarray}
with $\Delta=r^2+a^2-2r_g r$,  $\Sigma=r^2+a^2\cos^2\theta$,
$A=(r^2+a^2)^2-a^2\Delta\sin^2\theta$, and $r_g=GM/c^2$ denotes the gravitational radius.
The parameter $a=J/M$ represents the specific 
angular momentum.  The determinant of the matrix $g_{\mu\nu}$ is given by $\sqrt{-g}=\Sigma\sin\theta$.
The angular velocity of the black hole is defined as $\omega_H=\omega(r=r_H)=\tilde{a}/2r_H$, where 
$\tilde{a}=a/r_g$ denotes the dimensionless spin parameter, and $r_H=r_g+\sqrt{r_g^2-a^2}$ is the radius of the horizon. 
Henceforth, all lengths are measured in units of $r_g$ and time in units of $r_g/c$, so we set $c=r_g=1$ unless explicitly 
stated otherwise.

To avoid the singularity on the horizon, we find it convenient to transform to the tortoise 
coordinate $\xi$, defined by $d\xi= (r^2+a^2)dr/\Delta$.  It is related to $r$ through:
\begin{equation}
\xi(r)=r+\frac{1}{\sqrt{1-\tilde{a}^2}}\left[r_+\ln\left(\frac{r}{r_+}-1\right)
-r_-\ln\left(\frac{r}{r_-}-1\right)\right],
\label{tortoise_cord}
\end{equation}
with $r_\pm=1\pm \sqrt{1-\tilde{a}^2}$.  Note that $\xi\rightarrow-\infty$ as $r\rightarrow r_H=r_+$.

\subsection{Gap electric field}
\begin{figure}[]
\centering
\centerline{\includegraphics[width=8cm]{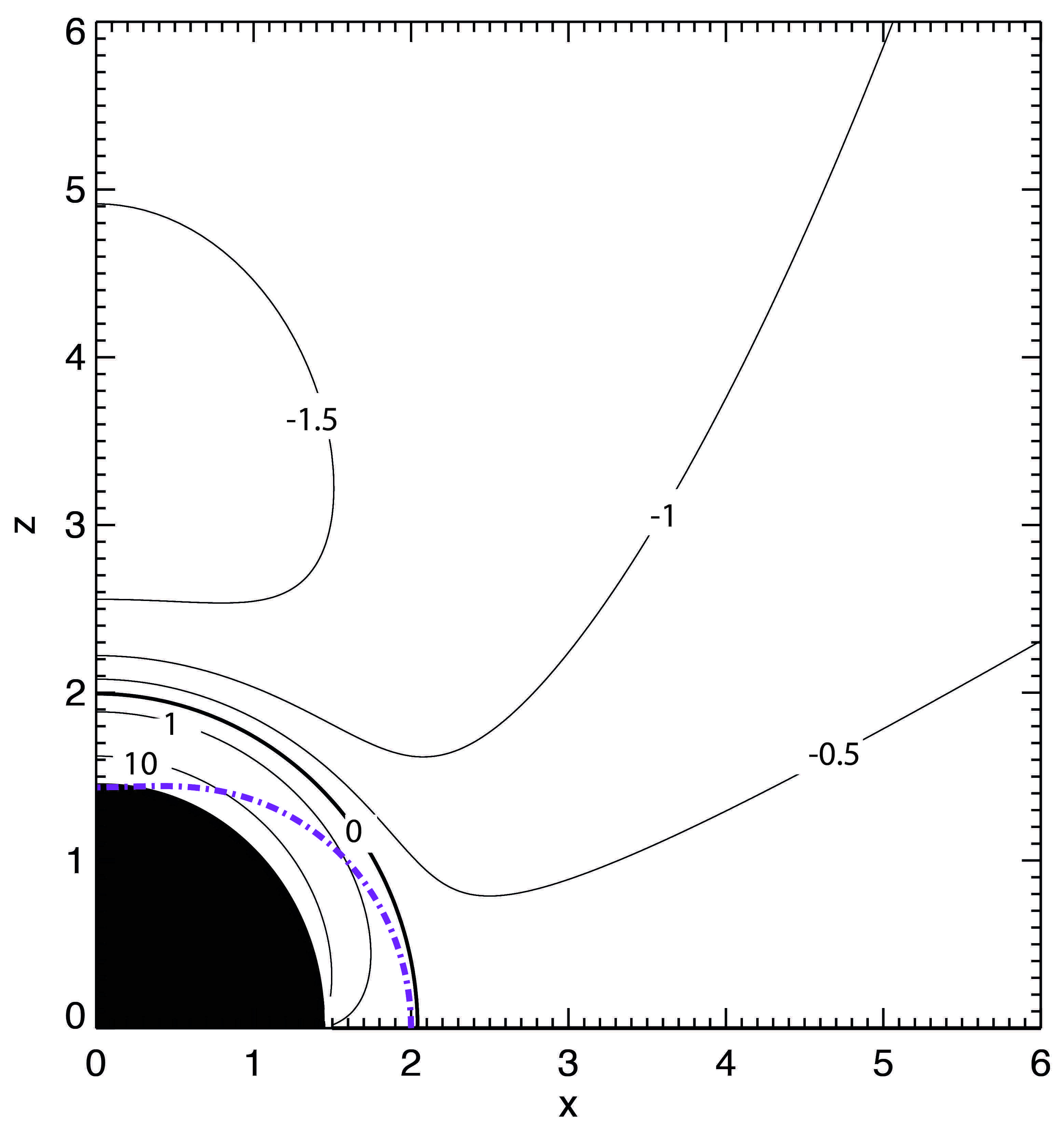}} 
\caption{Contours of the GJ charge density (solid lines), for $\tilde{a}=0.9$, $\Omega=0.5\omega_H$.   The numbers that label the curves are values of 
$\rho_{GJ}(r,\theta)/\rho_0$, where $\rho_0$ is the fiducial value defined below Equation (\ref{rho_GJ_1D_split}).   The thick solid line
corresponds to the null surface $r_c(\theta)$, on which $\rho_{GJ}=0$.
The black circle delineates the interior of the black hole, and the purple dashed-dotted line marks the static surface. }
\label{fig:rhogj}
\end{figure}

We implicitly assume that the gap forms a small perturbation in the force-free magnetosphere, in the sense that the 
potential drop across the gap is much smaller than the full vacuum potential.   We can then ignore the variation in $\Omega$ in the gap and,
for every magnetic flux surface, define the electric field in the corotating frame as $F^\prime_{\mu t}=F_{\mu t}+\Omega F_{\mu\varphi}$.
In general it satisfies Equation \ref{Gauss_app_rot}, with the GJ density defined explicitly in Equation \ref{rho_GJ_app}.   In order to compute
the gap structure in our formalizm, the magnetic field geometry needs to be specified.   In what follows we adopt a split monopole geometry,
defined by $A_\varphi=B_H\sqrt{A_H}(1-\cos\theta)$, where $B_H=10^8 B_{8}$ G denotes the strength of the magnetic field on the horizon, and 
$A_H\equiv A(r=r_H)=(r^2_H+a^2)^2=4(1+\sqrt{1-\tilde{a}^2})^2$ in our units ($4r_g^2r_H^2$ in full units).  With this choice $F_{r\varphi}=0$ and $F_{\theta\varphi}=B_H\sqrt{A_H}\sin\theta$. 
Note that in the ZAMO frame the radial magnetic field is given by $B_r=F_{\theta\varphi}/\sqrt{A}\sin\theta = B_H\sqrt{A_H/A}$, 
and the non-corotating electric field by $E^\prime_r=\sqrt{A} F^\prime_{rt}/\Sigma$.   We find it convenient to use the electric flux function $\Phi_E=\sqrt{A} E^\prime_r$
(which is essentially the electric flux per solid angle, as measured in the ZAMO frame).  
Then, Gauss' law, Eq. (\ref{Gauss_app_rot}), reduces to (see Eq. (\ref{Gauss_app_split}))
\begin{equation}
\partial_\xi \Phi_E=\frac{4\pi \Sigma\Delta}{r^2+\tilde{a}^2} (\rho_e-\rho_{GJ}),\label{1D_Poiss}\\
\end{equation}
with the GJ density given by
\begin{equation}
\rho_{GJ}=\frac{B_H\sqrt{A_H}}{4\pi \sqrt{-g}}\partial_\theta \left[\frac{\sin^2\theta}{\alpha^2}(\omega-\Omega)\right].
\label{rho_GJ_1D_split}
\end{equation}
Note that $\Sigma \Delta \rho_{GJ}$ is finite on the horizon.   Contours of $\rho_{GJ}(r,\theta)$ are exhibited in Figure \ref{fig:rhogj} for $\Omega=0.5\omega_H$ and $\tilde{a}=0.9$.  As seen,  it vanishes on the null surface denoted here by $r_c(\theta)$, located roughly (but not exactly) where $\Omega=\omega$.
In what follows the charge density and electric flux are normalized to the fiducial values $\rho_0=B_H\omega_H\sqrt{A_H}/(2\pi cr_g^2)=B_H\tilde{a}/2\pi r_g$
and $\Phi_o=\rho_0 r_g^3$, respectively, densities are measured in units of $n_0=\rho_0/e$, and angular velocities are measured in units of $\omega_H$.    
With the convention ${\bf \Omega}\cdot{\bf B}>0$ adopted below the electric field in the gap is negative, $\Phi_E<0$.

\subsection{Plasma dynamics}

We adopt a treatment in which the plasma in the gap is modelled as a two-component fluid, consisting of electrons and positrons
with proper number densities $n_-$ and $n_+$, respectively, and 4-velocities $u^\mu_\pm=(u_\pm^t,u_\pm^r,0,0)$.   
In a ZAMO frame the velocity components are given by $u_\pm=\sqrt{g_{rr}}u_\pm^r$, $\gamma_\pm =\alpha u_\pm^t$, $v_\pm=u_\pm/\gamma_\pm$.
We define the radial fluxes, $N^r_\pm=\Sigma n_\pm u^r_\pm$, measured in units of $r_g^2 n_0 c$. 
The continuity equation for each species can then be expressed as (see appendix \ref{app:fluid}),
\begin{equation}
\partial_\xi N_\pm^r= \frac{\Sigma \Delta}{2(r^2+\tilde{a}^2)} Q,
\label{eq:cont}
\end{equation}
where $Q$ is the net pair production rate per unit volume, measured in units of $n_{0}c/r_g$, and is the same for electrons and positrons by 
virtue of charge conservation.  It is readily seen that the difference $N_0^r=N^r_+-N_-^r$ is conserved along magnetic flux tubes.   This conserved  
quantity is simply the electric current per solid angle per unit charge flowing along magnetic flux tube, viz.,  $N_0^r= \Sigma j^r/e$,
where $j^r=e(n_+u^r_+-n_- u_-^r)$ is the radial component of the electric 4-current density, which is determined by the global magnetospheric 
structure.  The evaluation of $N_0^r$ requires proper account of the coupling between the gap and the global magnetosphere, which 
is beyond the scope of our analysis, and in our model it is treated as a free parameter.   As will be shown below, it 
affects the gap structure.   The normalized charge density, $\rho_e= j^t/\rho_0=e(n_+ u_+^t - n_-u^t_-)/\rho_0$, can be expressed in terms of the
electron and positron  fluxes as, 
\begin{equation}
\rho_e=\frac{\sqrt{A}}{\Sigma\Delta}\left(\frac{N^r_+}{v_+}-\frac{N^r_-}{v_-}\right).
\end{equation}

With the convention ${\bf \Omega}\cdot{\bf B}>0$ ($\Phi_E<0$) electrons accelerate outwards, $v_->0$, and positrons inwards, $v_+<0$.  The equations of motion 
of the pair fluids can be expressed as (see appendix \ref{app:fluid} for details),
\begin{equation}
\frac{d\gamma_\pm}{d\xi}=- \gamma_\pm\partial_\xi\ln\alpha \pm \frac{\alpha}{r^2+\tilde{a}^2}\left( \eta_E\,\Phi_E -\sqrt{A} s^t_\pm \right),
\label{eq_of_mot}
\end{equation}
with
\begin{equation}
\eta_E=\frac{eB_H\sqrt{A_H}\omega_H}{2\pi m_e c^3}
=1.4\times 10^9 \tilde{a} B_8 m.
\label{eta_E}
\end{equation}
The first term on the right hand side of Equation (\ref{eq_of_mot}) accounts for the gravitational redshift, the second term
for energy gain due to acceleration in the gap electric field, and the third term for the sum of curvature and inverse 
Compton losses, $s_\pm^t=s^t_{\pm,cur}+s^t_{\pm,IC}$, derived explicitly below.   As will be shown below, in practice
the Lorentz factors $\gamma_\pm$ equal their saturation values, at which energy gain is compensated by redshift effects and radiative losses
almost everywhere in the gap. 

\subsection{Gamma-ray emission and pair production}
We suppose that the gap is exposed to emission of soft photons by the accretion flow, 
from a putative source of size $R_s=\tilde{R}_s r_g$ and luminosity $L_s=l_s L_{Edd}$.  
For simplicity, we assume that the intensity 
of the seed radiation in the gap is isotropic with a power law spectrum: 
$I_s(x^\mu,\nu_s,{\bf \Omega}_s)=I_0(\epsilon_s/\epsilon_{s,min})^{-p}$,  $\epsilon_{s,min}<\epsilon_s<\epsilon_{s,max}$,
where $\epsilon_s=h\nu_s/m_ec^2$ is the dimensionless photon energy and $p>1$.  
The assumption that $I_s$ is isotropic is reasonable, except perhaps very near the horizon, 
since the size $R_s$ of the radiation source is typically much larger than the gap dimensions.
The number density of seed photons is given by 
\begin{equation}
n_s=\frac{4\pi}{c}\int \frac{I_s}{h\nu_s}d\nu_s=\frac{4\pi I_0}{h c}\frac{(1-\epsilon_{s,min}^p/\epsilon_{s,max}^p)}{p}
\simeq \frac{4\pi I_0}{h c}.
\end{equation}
We find it convenient to define a fiducial optical depth:  
\begin{equation}\label{fiducial-tau}
\tau_0=\sigma_T r_g \frac{4\pi I_0}{hc}=A_s\frac{4 m_p}{m_e}\frac{l_s}{\tilde{R}_s^2 \epsilon_{s,min}},
\end{equation}
where $A_s=(p-1)/[1-(\epsilon_{s,min}/\epsilon_{s,max})^{p-1}]\sim 1$.  It roughly gives the scaling of the IC and pair production opacities.   Typically $\tilde{R} < 10^2$,
so that a large opacity is anticipated when $l_s>\epsilon_{s,min}$.  

As shown below, the terminal Lorentz factor of the pairs in the gap is extremely high.  Thus, their
emission is highly beamed along their direction of motion.   
Let $I_\gamma(r,\epsilon_\gamma, \mu_\gamma)$ denotes the intensity of gamma-rays emitted by the pairs at radius $r$, 
in direction $\mu_\gamma=\cos\theta_\gamma=\hat{r}\cdot\hat{\Omega}_\gamma$ 
and energy $\epsilon_\gamma=h\nu_\gamma/m_ec^2$.  Under the beaming approximation we have: 
\begin{equation}
I_\gamma(r,\epsilon_\gamma, \mu_\gamma)=I_\gamma^+(r,\epsilon_\gamma)\delta(\mu_\gamma+1)
+I_\gamma^-(r,\epsilon_\gamma)\delta(\mu_\gamma-1), 
\label{beam_gamma}
\end{equation}
here $I^-_\gamma$ denotes the intensity emitted by electrons and $I_\gamma^+$ by positrons.
The beamed intensities satisfy the radiative transfer equations
\begin{eqnarray}
\frac{1}{\sqrt{A}}\frac{d}{d\xi}(\sqrt{A}I^\pm_\gamma )=\pm\frac{\alpha\sqrt{A}}{r^2+\tilde{a}^2}\left( \kappa_{pp}I^\pm_\gamma -  j_{\gamma}^\pm\right),
\label{transfer_eq}
\end{eqnarray}
neglecting redshift effects (see appendix \ref{app:radiation} for details), where the emissivity is the sum of curvature and IC emissions, 
$j^\pm_\gamma=j^\pm_{IC} + j_{cur}^\pm$.  The absorption coefficient
$\kappa_{pp}$ and the emissivities $j^\pm_{IC}$ and $j^\pm_{cur}$ are computed in the ZAMO frame.
To render this equation dimensionless, we normalize intensities by $h c n_0$, emissivities 
by $hc n_0/r_g$, and opacities by $1/r_g$.

\subsubsection{Curvature emission}
The normalized curvature emissivity is given by \cite{RL79}
\begin{equation} 
j_{cur}^\pm (r,\epsilon_\gamma)=\frac{\sqrt{3}\alpha_f  n_\pm \gamma^2_\pm}{2\pi R_c} F(\epsilon/\epsilon_c),
\label{j_cur_1}
\end{equation}
where $R_c$ denotes the  curvature radius of magnetic field lines (in units of $r_g$),  $\alpha_f=e^2/\hbar c$ is the fine structure constant, $F(x)$ is 
the usual synchrotron function, and
\begin{equation} 
\epsilon_c=\frac{2\pi \lambda_c}{r_g} \frac{\gamma_\pm^3}{R_c}\simeq 10^{-15} \frac{\gamma_\pm^3}{m\, R_c},
\label{eps_c_1}
\end{equation}
here $\lambda_c=\hbar/m_ec$ denotes the Compton wavelength of the electron.   The curvature radius is a free parameter in our model.  In 
the numerical calculations presented below we adopted $R_c = 1$.
Finally, the curvature loss term is given by
\begin{equation}
s^t_{\pm,cur}= -10^{-18}\frac{\gamma_\pm^4}{m R_c^2}.
\end{equation}

\subsubsection{Inverse Compton emission}
The normalized IC emissivity, computed in appendix \ref{app:radiation} using the full Klein-Nishina (KN) cross-section, can be expressed in the ZAMO frame in terms of the fiducial optical depth $\tau_0$ as:
\begin{eqnarray} \label{j_IC_1}
j^\pm_{IC}(r,\epsilon_\gamma)&=& \frac{\tau_0 n_\pm\gamma_\pm}{6\pi} \left[\frac{4\gamma_\pm\epsilon_{s,min}(\gamma_\pm-\epsilon_\gamma)}{\epsilon_\gamma+4\gamma_\pm \epsilon_{s,min}(\gamma_\pm-\epsilon_\gamma)}\right]^{p}\\
&\times&  \left[\frac{\epsilon_\gamma}{\epsilon_\gamma+4\gamma_\pm\epsilon_{s,min}(\gamma_\pm-\epsilon_\gamma)}  \right].\nonumber
\end{eqnarray}
The corresponding drag (energy loss) terms for the pairs are given formally by
\begin{equation}
s^t_{\pm,IC}(r)=- \frac{2\pi}{n_\pm \gamma_\pm}\int_0^{\gamma_\pm} j^\pm_{IC}(r,\epsilon_\gamma) d\epsilon_\gamma.
\end{equation}
Within the beaming approximation invoked here, transition from the Thomson to the KN regime occurs at a Lorentz factor $\gamma_{KN}=1/4\epsilon_{s,min}$.  
In the Thomson limit, where $\gamma/\gamma_{KN}=4\gamma_\pm\epsilon_{s,min}<<1$, the above expressions reduce to
\begin{equation}
j^\pm_{IC}(r,\epsilon_\gamma)\simeq \frac{\tau_0 n_\pm\gamma_\pm}{6\pi} \left(\frac{\epsilon_\gamma}{4\gamma_\pm^2\epsilon_{s,min}} \right)^{-p},
\end{equation}
with $4\gamma_\pm^2\epsilon_{s,min}\le\epsilon_\gamma $, and 
\begin{equation}
s^t_{\pm,IC}(r)\simeq -\frac{4 \tau_0}{3(p-1)} \gamma_\pm^2\epsilon_{s,min},
\end{equation}
whereas in the KN limit, $\gamma_\pm/\gamma_{KN}>>1$, we have 
\begin{equation}
j^\pm_{IC}(r,\epsilon_\gamma)\simeq \frac{\tau_0 n_\pm\gamma_\pm}{6\pi} \left(\frac{\epsilon_\gamma}{4\gamma_\pm^2\epsilon_{s,min}} \right),
\end{equation}
and 
\begin{equation}
s^t_{\pm,IC}(r)\simeq -\frac{\tau_0}{24\, \epsilon_{s,min}}.
\end{equation}

\subsubsection{Pair production}

Under the assumption that the seed photon intensity is isotropic, the normalized pair production opacity simplifies to
\begin{eqnarray}
\kappa_{pp}(r,\epsilon_\gamma)&=&\frac{\tau_0}{2 } \int_{\epsilon_{th}}^{\epsilon_{s,max}}
d\ln \epsilon_s\left(\frac{\epsilon_s}{\epsilon_{s,min}}\right)^{-p}\\
&\times& \int_{-1}^{\mu_{max}}d\mu(1-\mu)\sigma_{\gamma\gamma}\nonumber
\end{eqnarray}
where $\sigma_{\gamma\gamma}$ is the full pair creation cross-section (in units of $\sigma_T$) given in Ref~\cite{GS67}, $\epsilon_{th}={\rm max}(\epsilon_{s,min},\, \epsilon_\gamma^{-1})$ and $\mu_{max}=1-2/(\epsilon_s\epsilon_\gamma)$ from the threshold condition.   In the Thomson limit, that is, $\epsilon_\gamma\epsilon_{s,min} <1$, it is given, to a good approximation, by \cite{BL95} 
\begin{equation}
\kappa_{pp}(r,\epsilon_\gamma)=\frac{3 \tau_0}{2 } A_p(\epsilon_{s,min} \,\epsilon_\gamma)^p,
\label{kpp_Thom}
\end{equation}
where $A_p$ is a number that depends on the spectral index $p$, and is plotted in Figure 1 of Ref~\cite{BL95}.  
It equals roughly $0.2$ for $p=1$ and $0.1$ for $p=2$.   In the KN limit, $\epsilon_\gamma\epsilon_{s,min} >> 1$, we can use the approximation 
$\int_{-1}^{\mu_{max}}(1-\mu)\sigma_{\gamma\gamma} d\mu \simeq 3\ln(\epsilon_\gamma\epsilon_{s})/2\epsilon_\gamma\epsilon_{s}$, to obtain
\begin{equation}
\kappa_{pp}(r,\epsilon_\gamma)=\frac{3 \tau_0}{4(p+1) } \frac{\ln(\epsilon_\gamma\epsilon_{s,min})}{\epsilon_\gamma\epsilon_{s,min}}.
\label{kpp_KN}
\end{equation}
For the parameter regime considered here we find $\ln(\epsilon_\gamma\epsilon_{s,min})/2(p+1)\simeq 1$.  Thus, to a good approximation we can 
use the simple extrapolation:
\begin{equation}
\kappa_{pp}(r,\epsilon_\gamma)=\frac{3\tau_0}{2 }\frac{A_p(\epsilon_{s,min} \,\epsilon_\gamma)^p}{1+A_p(\epsilon_{s,min}\, \epsilon_\gamma)^{p+1}}.
\label{kpp_full}
\end{equation}
The net specific pair production rate can be expressed as
\begin{eqnarray}\label{Q_1}
Q&=&2\pi  \int d\ln \epsilon_\gamma \int_{-1}^{1}d\mu_\gamma \kappa_{pp} I_\gamma(r,\epsilon_\gamma,\mu_\gamma)\\
&=&3\pi \tau_0 \int \frac{A_p(\epsilon_\gamma\epsilon_{s,min})^{p}}{1+A_p(\epsilon_\gamma\epsilon_{s,min})^{p+1}}
\left(I^+_\gamma+I_\gamma^-\right)d\ln \epsilon_\gamma.\nonumber
\end{eqnarray}

\subsection{Boundary conditions}

The outer and inner gap boundaries are treated as free boundaries.  Their location is determined 
by two parameters; the global current $N_0^r$ and the fiducial optical depth $\tau_0$.  
In the steady gap model it is implicitly assumed that beyond the gap boundaries the field aligned electric field vanishes.  Thus, it must satisfy the boundary conditions
\begin{equation}
\Phi_E(r_{in})=\Phi_E(r_{out})=0.
\label{bc-1}
\end{equation}
We further assume that pairs and photons are not injected into the gap across either boundary.  
This implies 
\begin{eqnarray}
N^r_{+}(r_{out})=N^r_{-}(r_{in})=0, \nonumber \\
N^r_{-}(r_{out})=- N^r_{+}(r_{in})=- N_0^r.
\label{bc-2}
\end{eqnarray}
Likewise, since no gamma-rays are incident into the gap through its boudaries,
\begin{equation}
I_\gamma^+(r_{out},\epsilon_{\gamma})= I_\gamma^{-}(r_{rin},\epsilon_{\gamma})=0.
\label{bc-3}
\end{equation}
The Lorentz factors of the electron and positron beams formally satisfy $\gamma_-(r_{in})=\gamma_{+}(r_{out})=1$.   However, 
since practically they reach their saturation level instantaneously, we find that the solution is highly insensitive to the exact values taken at the boundary, as long as 
they are much smaller than the maximum values.

\begin{figure*}[]
\centering
\centerline{\includegraphics[width=8cm]{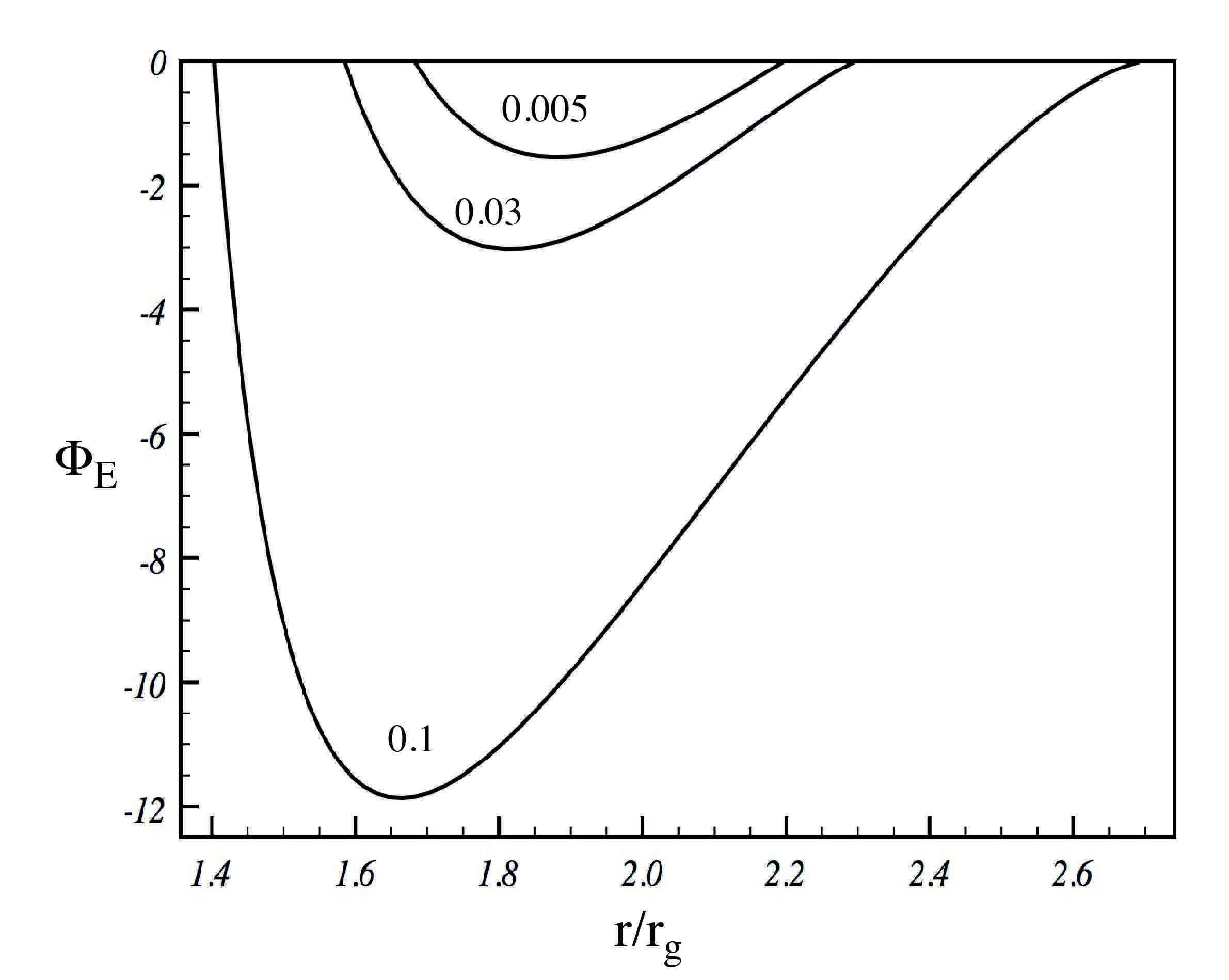}  \includegraphics[width=8cm]{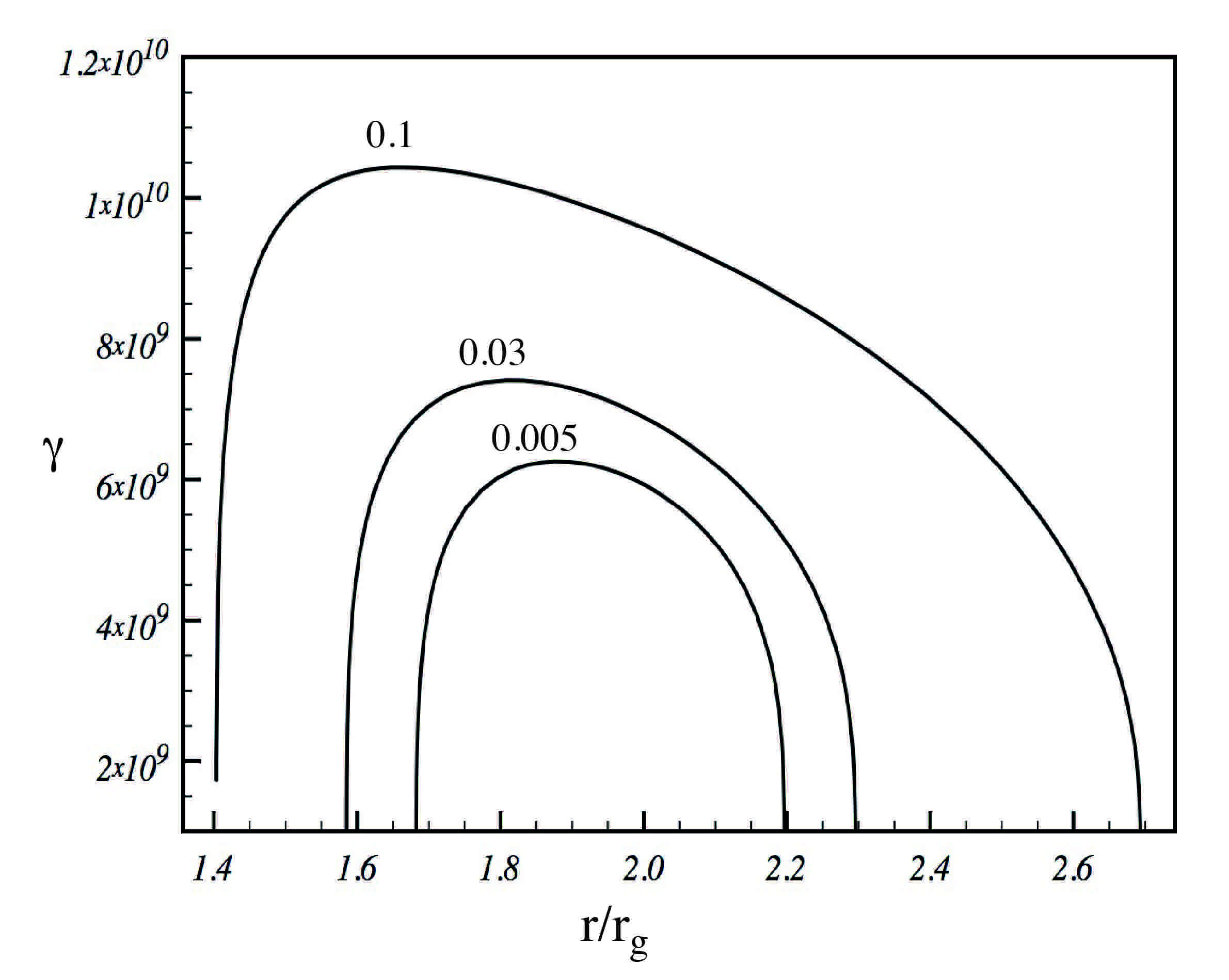}}
\centerline{ \includegraphics[width=8cm]{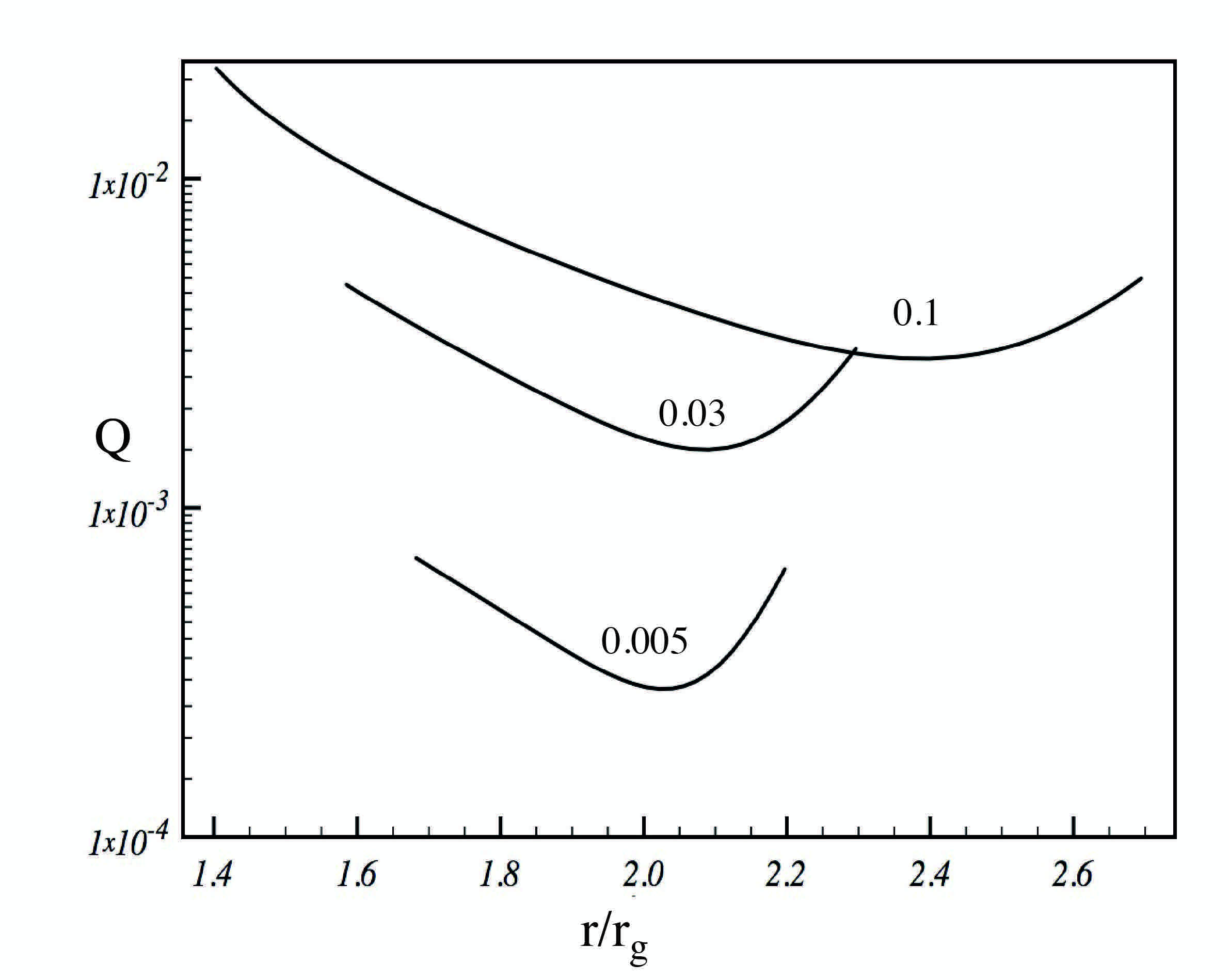}  \includegraphics[width=8cm]{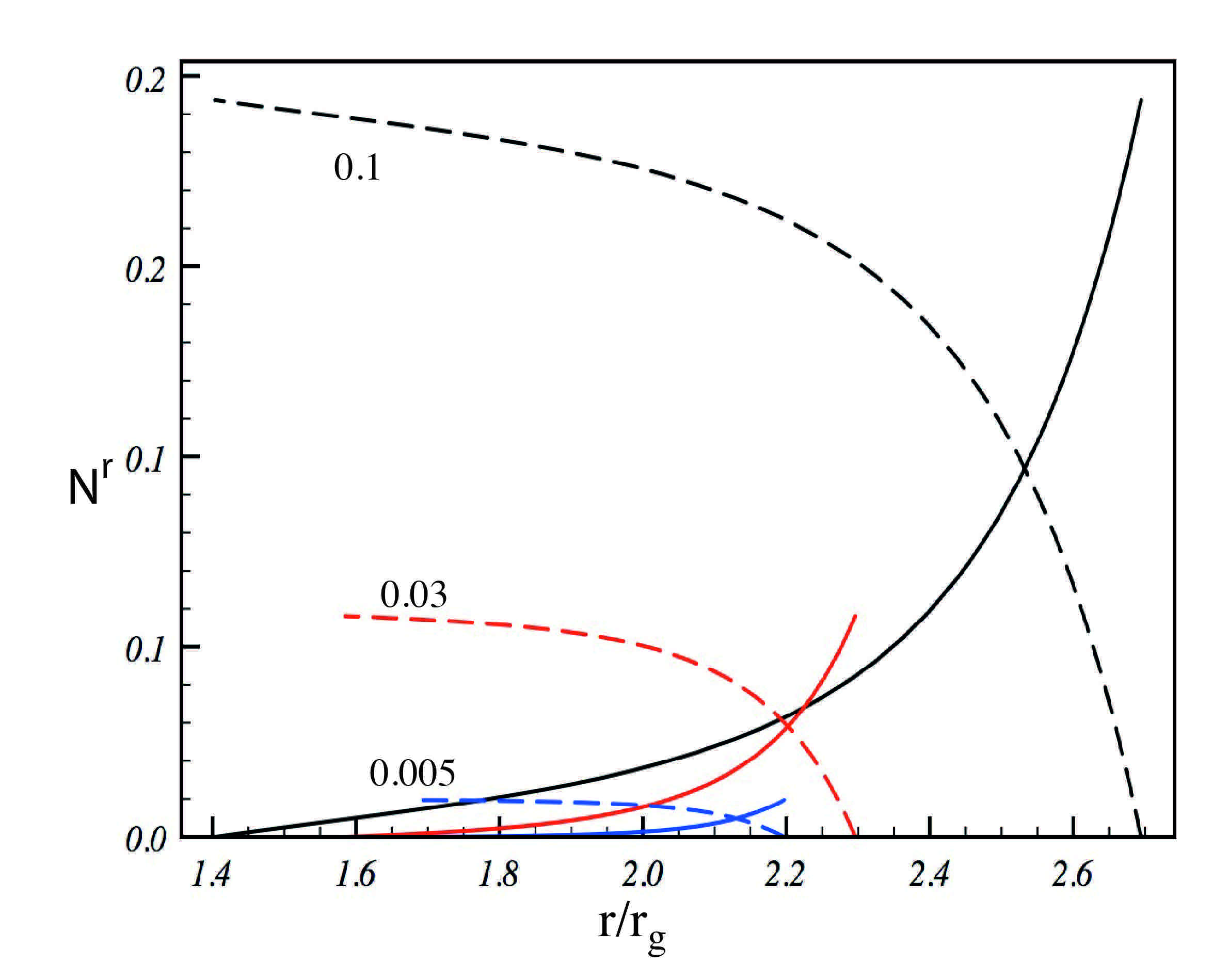}}
\caption{Solutions of a steady gap for different values of the normalized magnetospheric current, $2\pi j^r_c/(\Omega B_H\cos\theta)$ (indicated
by the numbers that label the curves).  The parameters used in all cases shown are $\theta=30^\circ$,  $M=10^9 M_\odot$, $B_8=10^{-6}$, $\epsilon_{s,min}=10^{-8}$.
Shown are  the profiles of the electric flux (upper left panel), Lorentz factor of the pairs (upper right panel), pair creation rate (lower left panel) and 
particle fluxes (lower right panel). The solid lines in the lower right panel correspond to the electron flux $N^r_{-}$ and the dashed lines to the positron flux  $N^r_{+}$.}
\label{fig:profiles}
\end{figure*}

\section{\label{sec:results}Steady gap solutions and forbidden regimes}

Equations (\ref{1D_Poiss})-(\ref{eq_of_mot}), (\ref{transfer_eq}), (\ref{j_cur_1}), (\ref{j_IC_1}), (\ref{kpp_full}) and (\ref{Q_1}), subject to the
boundary conditions (\ref{bc-1})-(\ref{bc-3}) form a complete set that governs the structure and spectrum of the steady gap for a given choice
of the input parameters $N_0^r$ and $\tau_0$ (if a solution exists).   The location of the outer boundary $r_{out}$ is constrained  to exceed a minimum 
value by the condition $|\rho_e(r_{out})|< |\rho_{GJ}(r_{out})|$.   To obtain a solution, we integrate the equations iteratively, changing the locations
of the inner and outer boundaries, $r_{in}$ and $r_{out}$, in each iteration, until all boundary conditions are satisfied.   In each iteration 
we first guess a value for $r_{out}$, and then integrate the equations inwards starting at $r_{out}$ until $\Phi_E$ vanishes (provided
it is outside the horizon).   We then check the values of $N^r_{-}$ and $I_\gamma^{-}$ there, and if nonzero change the location of $r_{out}$ accordingly 
for the next iteration. The process is repeated until the desired solution is obtained.  

Examples are exhibited in Figure \ref{fig:profiles}, where profiles of the electric flux, Lorentz factor, pair fluxes and specific pair production rate, 
computed for a prototypical supermassive BH accreting in the RIAF regime ($\dot{m}\simeq10^{-4}$), are plotted for different values of the 
magnetospheric current, here 
represented in terms of the current density at the null 
surface, $j_c=e N^r_0/\Sigma_c$ (normalized by the fiducial current $\Omega B_H \cos\theta/2\pi$), 
where $\Sigma_c\equiv \Sigma(r_c)$ is the value of $\Sigma(r)$ at the null surface $r_c(\theta)$.  
Each case shown corresponds to a specific value of $\tau_0$.
Similar solutions were obtained for parameters typical to stellar BHs. 
As seen, the gap shrinks as the magnetospheric current $j_c$ (or equivalently the flux $N_0^r$) is reduced, as expected.   It is 
also seen that unless the magnetospheric current is unlikely weak, the gap width is not much smaller than  the horizon scale. 
Since in a stationary gap the pair multiplicity cannot largely exceed unity (see bottom right panel in Fig. \ref{fig:profiles}), this implies that $\tau_0$ (and hence $L_s$) must
also be small, as shown next.

Much insight can be gained into the behaviour of the gap by employing crude estimates that allow analytic  derivation of the pair production 
rate and the closure condition.   Below, we adopt such a treatment to map the parameter regime in which local, steady gap solutions exist.

\begin{figure*}[]
\centering
\centerline{\includegraphics[width=16cm]{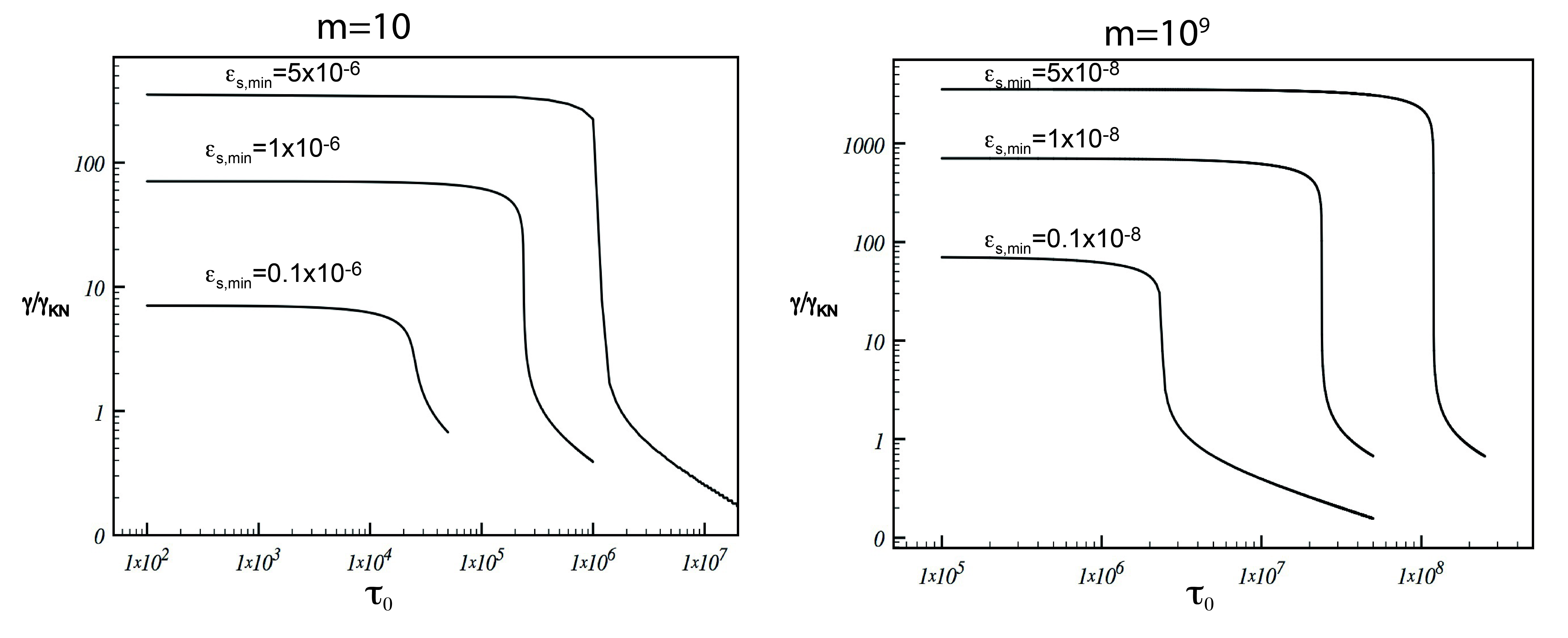}}  
\caption{Dependence of the terminal Lorentz factor $\gamma$ (normalized to $\gamma_{KN}=1/4\epsilon_{s,min}$) on the fiducial depth $\tau_0$, for different values
of  $\epsilon_{s,min}$ (indicated by the numbers that label the curves).  The left panel corresponds to
a stellar black hole with $m=10$ and $B_8=1$, and the right panel to a supermassive black hole with $m=10^9$ and $B_8=10^{-4}$.   The sharp decline 
delineates the transition from curvature dominated to IC dominated drag.}
\label{fig:gsat}
\end{figure*}

The Lorentz factor of accelerating pairs is limited by the saturation value at which the radiation drag (due to curvature and IC emission) balances the 
electric force acting on the particles within the gap.
It formally obtained by setting the right hand side of Eq. (\ref{eq_of_mot}) to zero.     Neglecting gravity (which is important only very near the horizon)
we find that the acceleration length is roughly $l_{acc}
\simeq 10^{-2} m^{-1/2}R_c^{1/2} (B_8| E_r^\prime|)^{-3/4}$,
so that practically the Lorentz factor is determined by the saturation condition in the entire gap region.   
The dependence of $\gamma/\gamma_{KN}$, the saturated Lorentz factor normalized by  $\gamma_{KN}\equiv 1/4\epsilon_{s,min}$,  
on $\tau_0$ is displayed in Fig 3, for different values of the peak energy $\epsilon_{min}$.  
The transition from curvature dominated to IC dominated losses is clearly seen.  The value of $\tau_0$ at which the transition occurs depends
on the spectral peak $\epsilon_{min}$ through Klein-Nishina effects.
It can be estimated analytically from the saturation condition, whereby it is found
 that IC losses dominate the drag force when 
$\tau_0>10^4\,B_8\,m (\epsilon_{s,min}/10^{-6})$, as indeed seen in Figure \ref{fig:gsat}.  As argued below, for such high values of $\tau_0$ steady gap solutions do not exist for any reasonable choice of parameters, hence this regime is irrelevant for our analysis.     At smaller values of $\tau_0$ curvature losses dominate, and  the saturated Lorentz factor is:
\begin{equation}
\gamma_\pm  \simeq 5\times10^6 R_c^{1/4}(B_8|E^\prime_r|)^{1/4} m^{1/2},
\label{gam_sat}
\end{equation}
so that $\gamma_+=\gamma_-=\gamma$.  Under a broad range of conditions we find $|E^\prime_r|^{1/4}\sim1$.  Hence, for our fiducial stellar black hole, $m=10$, $B_8=1$, we expect $\gamma\sim 10^7$, whereas
for a fiducial blazar with  $m=10^9$ and $B_8=10^{-4}$ we have $\gamma\simeq 10^{10}$. Our detailed calculations confirm this. 
Under our beaming approximation, IC scattering is in the KN limit if (see Eq. \ref{j_IC_1})
\begin{equation}
\epsilon_{s,min} >\frac{1}{4\gamma}\simeq 5\times 10^{-8} R_c^{-1/4}(B_8|E^\prime_r|)^{-1/4} m^{-1/2}.
\end{equation}
This condition is satisfied essentially in all sources.  Consequently, we conclude that quite generally IC scattering of external radiation by pairs accelerated in the gap is in the KN regime (although sufficiently soft extension of the spectrum to energies below the peak may somewhat alter this conclusion).
The characteristic energy of curvature photons, Equation (\ref{eps_c_1}), can be expressed as
\begin{equation}
\epsilon_c\simeq 10^5 (B_8|E^\prime_r|)^{3/4} R_c^{-1/4}m^{1/2},
\label{eps_c_2}
\end{equation}
and it is seen that typically $\epsilon_c<<\gamma$.  Consequently, we expect two peaks in the high-energy spectrum emitted from the gap,
one due to IC scattering, at $\epsilon_\gamma\simeq\gamma$, and the other one due to curvature emission, at $\epsilon_\gamma \simeq 0.3 \epsilon_c$.
Equations (\ref{gam_sat}) and (\ref{eps_c_2}) imply that the separation between the peaks is independent of the black hole mass, but scales with the magnetic
field roughly as $B^{1/2}$.  The detailed calculations outlined in Ref~ \cite{Hir16} indeed confirm that the spectral energy distribution has a double peak structure with these scalings. 

Next, we provide estimates for the pair production opacity and the specific pair production rate in the gap.   Since IC scattering is in the KN regime, the characteristic energy 
of scattered photons is $\gamma$.    Thus, pair creation occurs also in the KN regime, whereby  Equation (\ref{kpp_KN}) applies:
\begin{eqnarray}
\kappa_{pp,IC}&=&\frac{3 \tau_0}{4(p+1) } \frac{\ln(\gamma \epsilon_{s,min})}{\gamma\epsilon_{s,min}}\sim 0.1 \tau_0\, R_c^{-1/4}\\
&\times&(B_8|E^\prime_r|)^{-1/4} m^{-1/2}\left(\frac{\epsilon_{s,min}}{10^{-6}}\right)^{-1},   \nonumber
\end{eqnarray}
for $p>1$.  Consequently, for $\tau_0>10 R_c^{1/4}$, $\kappa_{pp,IC}\simgt1$ for both stellar and supermassive black holes.  
At these energies the contribution of curvature emission is completely negligible (see Eq. (\ref{eps_c_2})), and 
the solution to the radiative transfer equation, Eq. (\ref{transfer_eq}), is approximately the IC source function, specifically $I^\pm_\gamma\simeq j^\pm_{IC}/\kappa_{pp}$.  Using Equation (\ref{Q_1}) and noting that in the ultra-relativistic limit $n_+\gamma_+ +n_-\gamma_- = -N^r_0/\sqrt{\Sigma\Delta}$, one obtains the contribution of IC scattered photons to the pair creation rate:
\begin{eqnarray}\label{Q_IC_4}
Q_{IC}(r)&\simeq& \frac{\tau_0}{12\gamma\epsilon_{min}}\frac{(-N^r_0)}{\sqrt{\Sigma\Delta}}\simeq 0.02 \tau_0 \frac{(-N^r_0)}{\sqrt{\Sigma\Delta}}\\
&\times&R_c^{-1/4}(B_8|E^\prime_r|)^{-1/4} m^{-1/2}\left(\frac{\epsilon_{s,min}}{10^{-6}}\right)^{-1}.\nonumber
\end{eqnarray}

The peak of curvature emission occurs 
at an energy of $\epsilon_\gamma=0.29 \epsilon_c$, for which $\epsilon_\gamma\epsilon_{s,min}  <<1$.   Thus, the interaction of curvature photons 
with the target radiation field is in the Thomson regime.  Choosing $p=2$ for illustration, Equation (\ref{kpp_Thom}) yields
\begin{eqnarray}\label{kpp,cur}
\kappa_{pp,cur}(\epsilon_\gamma)\simeq &10^{-4}&\, \tau_0~ m \,R_c^{-1/2}(B_8|E^\prime_r|)^{3/2}\\
&\times& \left(\frac{\epsilon_{s,min}}{10^{-6}}\right)^{2}\left(\frac{\epsilon_\gamma}{0.29\epsilon_c}\right)^{2}\nonumber
\end{eqnarray}
at energies $\epsilon_\gamma \simlt \epsilon_c$. Since for a steady gap $\tau_0<<10^4$, it implies $\kappa_{pp,cur}<<1$.
The calculation of $Q_{cur}$ is more involved than in the IC case, and we can only offer a rough analytic estimate of its average.
The details can be found in appendix \ref{sec:app-transfer}, where the following result for the average pair production rate is derived:
\begin{eqnarray}\label{Q_cur_av}
<Q_{cur}>&\simeq &2~ \tau_0~ (-N^r_0)R_c^{-5/4}(B_8|E^\prime_r|)^{7/4} \\
&\times& m^{3/2} \left(\frac{\epsilon_{s,min}}{10^{-6}}\right)^{2}
\frac{1}{<\sqrt{A}>}\int_{r_{in}}^{r_{out}}\frac{\sqrt{A} dr}{\Delta},\nonumber
\end{eqnarray}
here $<\sqrt{A}>$ is the average value of $\sqrt{A(r)}$ across the gap, defined explicitly below Equation (\ref{app_Ig-}), and is typically
in the range $3$ to $4.5$.
Equation (\ref{Q_cur_av})  may overestimate the local rate by a factor of a few.   From a comparison of Eqs. (\ref{Q_IC_4}) and (\ref{Q_cur_av}) 
we anticipate  the pair production to be dominated by IC photons
when $B_8 \simlt 0.1 m^{-1}(\epsilon_{min}/10^{-6})^{-3/2}R_c$.  This condition is roughly satisfied in RIAF sources with $\dot{m}<10^{-4}$,
assuming $R_c\simeq1$.   At larger accretion rates pair production is predominantly due curvature photons.

Finally, we derive a closure condition that defines a limit on the luminosity of the external radiation source, $l_s$, above
which steady gap solutions are forbidden.   For clarity of our analysis we include only the contribution of IC photons to the pair 
production rate, viz., $Q=Q_{IC}$.    Thus, the limit obtained from the closure condition derived below should be considered an absolute 
upper limit.  Additional production of pairs by curvature photons would merely enlarge the forbidden regime.
Integration of Equation (\ref{eq:cont}), subject to the boundary condition $N^r_{+}(r_{out})=0$,
yields $N^r_0=\int_{r_{in}}^{r_{out}}(\Sigma Q/2)dr$.   This last relation simply means that the pair multiplicity in the gap  is roughly unity.
Taking $Q=Q_{IC}$ in the latter expression and substituting Eq. (\ref{Q_IC_4}) yields
\begin{equation}
\tau_0=\frac{12 \gamma \epsilon_{min}}{H}\simeq 50
\frac{(R_cB_8|E^\prime_r|)^{1/4}m^{1/2}(\epsilon_{min}/10^{-6})}{H},
\end{equation}
where the factor $H=\int_{r_{in}}^{r_{out}}\sqrt{\Sigma/\Delta}~dr$  depends on the magnetospheric current $N^r_0$ through 
the gap boundaries $r_{in}$ and $r_{out}$, and is of order a few for the solutions shown in Figure \ref{fig:profiles}.   
It can become much smaller than unity for extremely small values of $N^r_0$, but we find such values unlikely.  For our fiducial sources 
the value of  $B_8^{1/4}m^{1/2}(\epsilon_{min}/10^{-6})$ is about 3 in case of a stellar BH and 
about 30 for a supermassive BH.   The maximum value of $|E^\prime_r|^{1/4}$ ranges between 0.8 and 1.3 in the solutions 
exhibited in Figure \ref{fig:profiles}.   
Consequently, the corresponding Eddington ratio, $l_s=1.3\times10^{-7} (\tilde{R}_s/30)^2(\epsilon_{min}/10^{-6})\tau_0$ (see
Eq. (\ref{fiducial-tau})), that allows stationary gap solutions must be very small.  Larger values would render the gap intermittent.

\begin{figure*}[]
\centering
\centerline{\includegraphics[width=8cm]{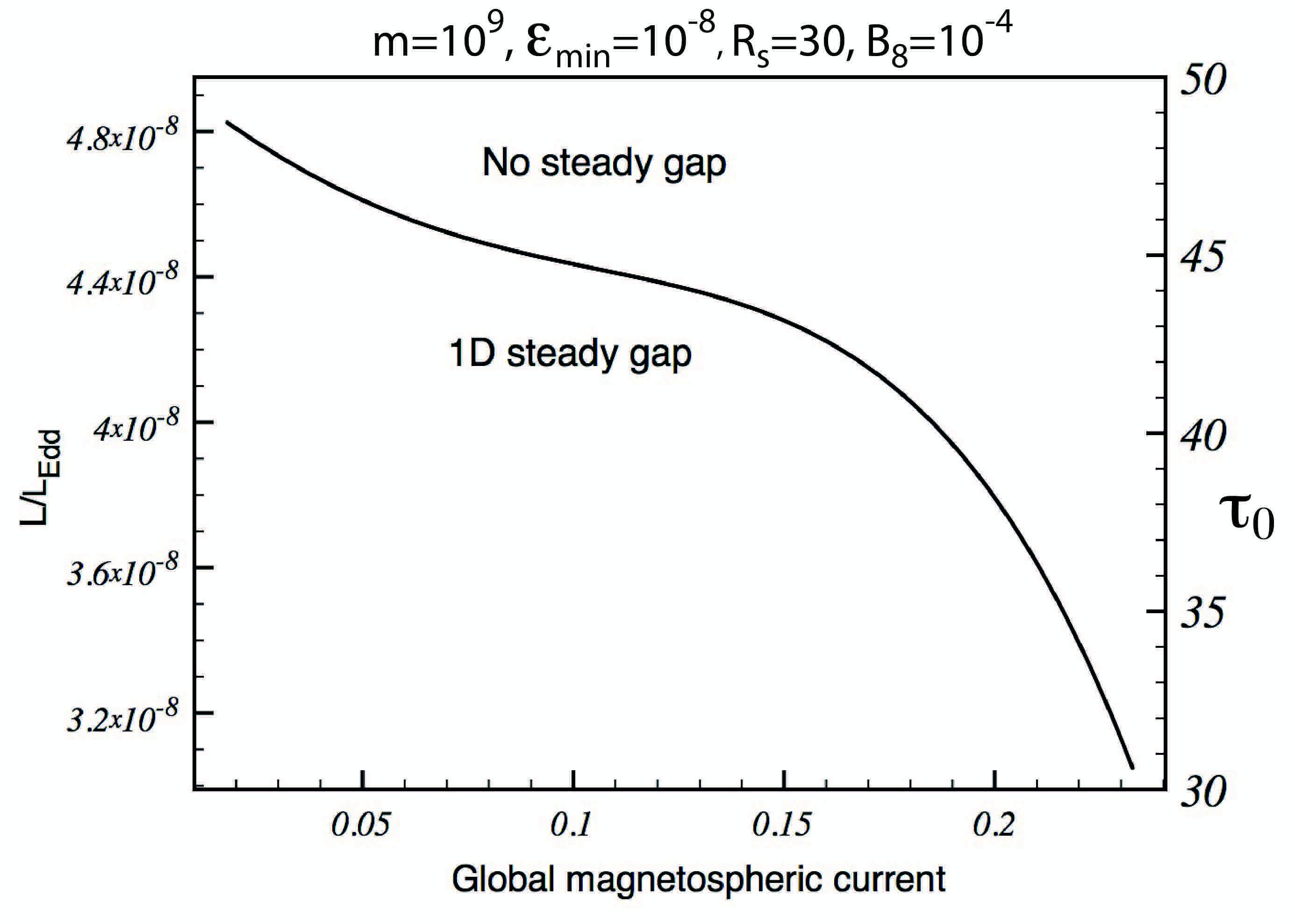}  \includegraphics[width=8cm]{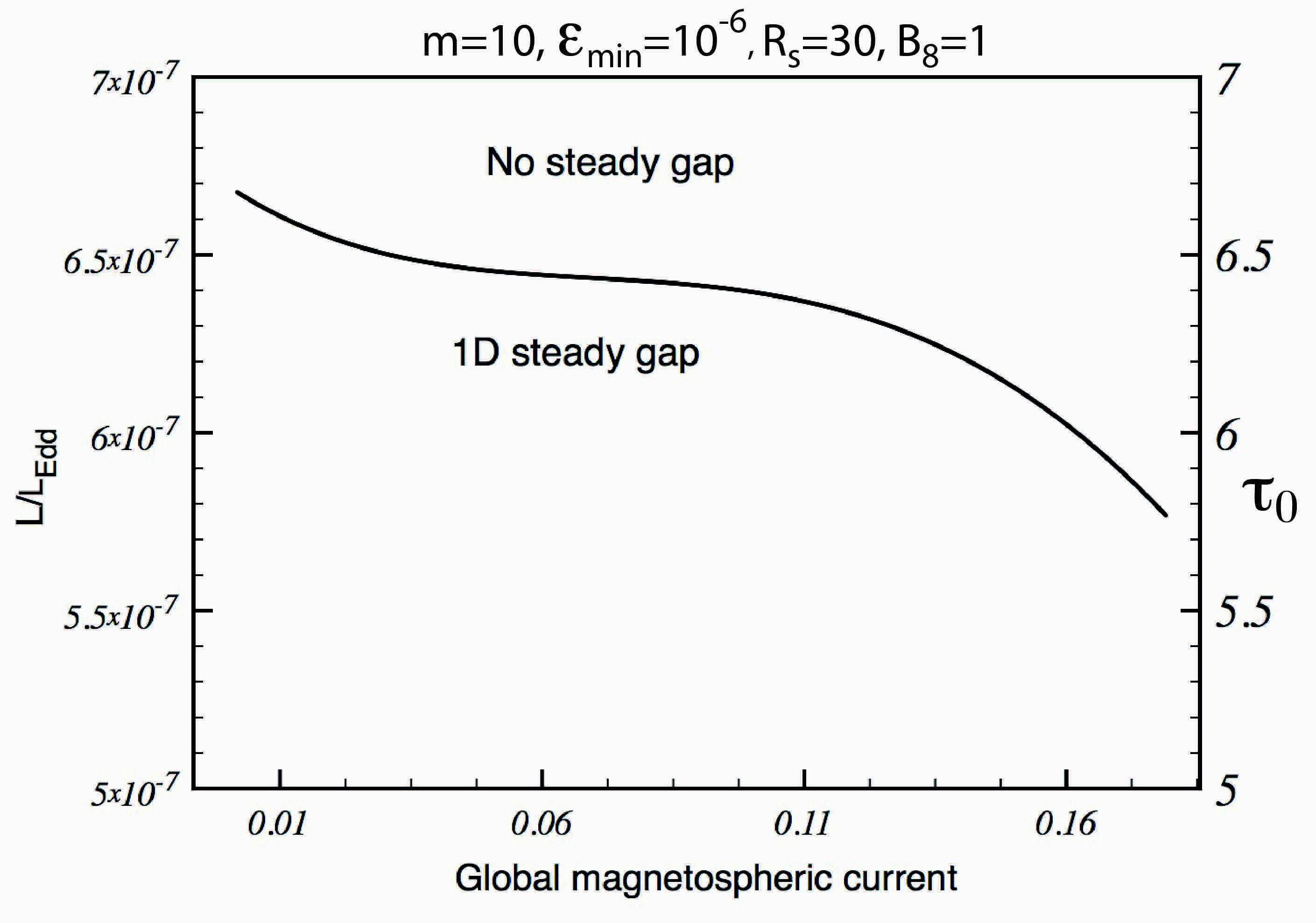}}
\caption{Maximum Eddington ratio below which local steady gap solutions exist, versus normalized magnetospheric current, $2\pi j^r_c/(\Omega B_H\cos\theta)$, 
for two fiducial sources, stellar BH and supermassive BH, characterized by the parameters indicated in the figure label.  The right axis shows 
the corresponding value of $\tau_0$.}
\label{fig:L-j}
\end{figure*}

Figure \ref{fig:L-j} shows the separation into forbidden and allowed regimes computed numerically using the full gap equations 
with $Q=Q_{IC}$.   The solid curve corresponds to the locus of solutions, each having the maximum value of $\tau_0$ above which no steady 
solutions exist.   For each choice of the magnetospheric current $N_0^r$ this maximum value is obtained by seeking the solution that 
satisfies $\rho_e(r_{out})=\rho_{GJ}(r_{out})$, or  equivalently $\sqrt{A(r_{out})}[\alpha(r_{out})]^2\rho_{GJ}(r_{out})=N_0^r$.
This solution defines the maximum luminosity $l_s$ at which a steady gap can still support  the current $N^r_0$.  At lower 
luminosities the gap widens ($r_{out}$ increases).  At larger luminosities it must become intermittent.

\section{\label{sec:remark} A remark on the global structure}
\begin{figure*}[]
\centering
\centerline{\includegraphics[width=5cm]{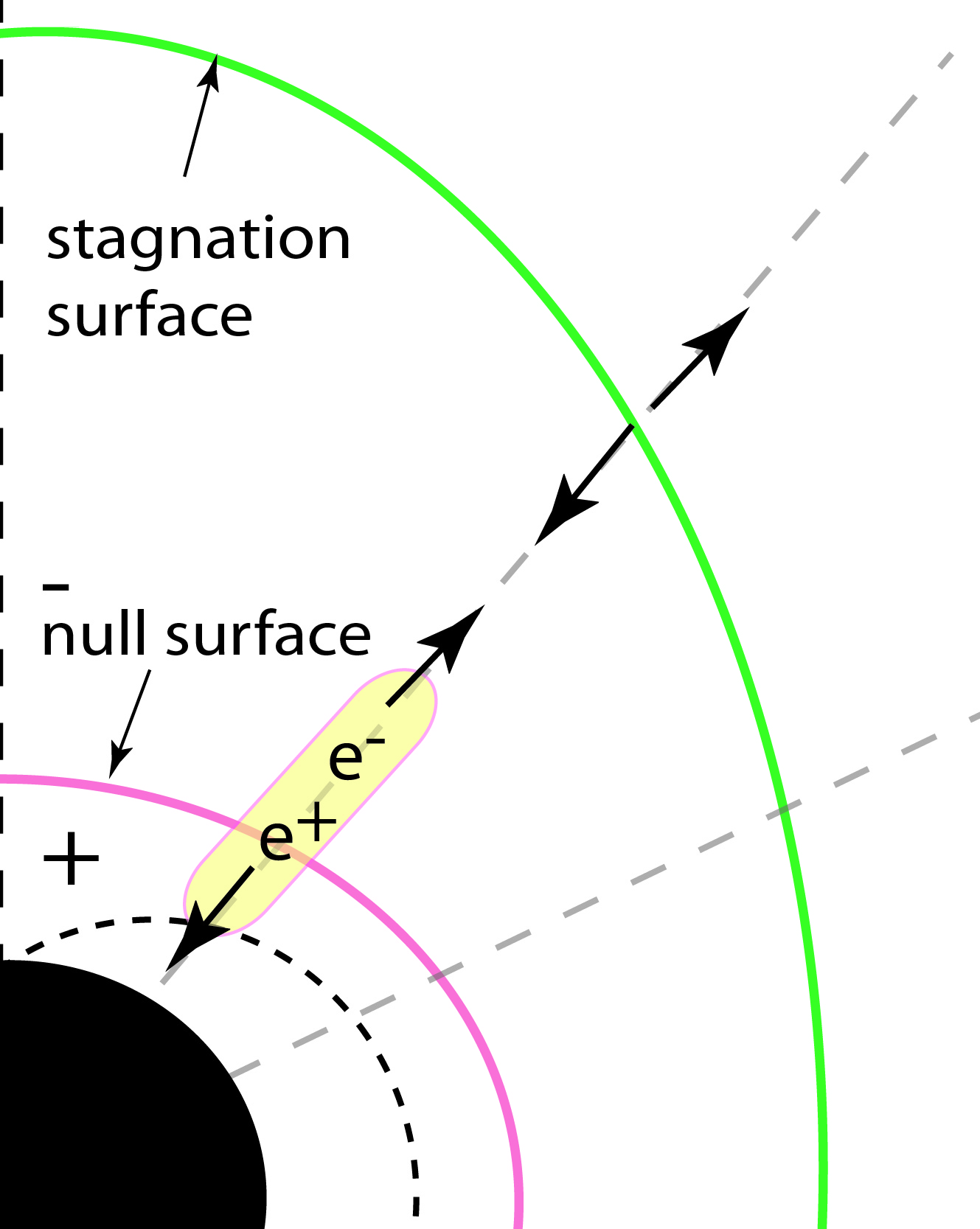}}
\caption{A sketch of the global structure.  The gap is represented by the yellowish stripe.  The thick arrows point to the direction 
of the electron (positron) beam leaving the outer (inner) gap boundary.  The two arrows that  emanate from the stagnation surface
show the flow directions of the plasma in the force-free region.}
\label{fig:global}
\end{figure*}

In this section we briefly comment on the relation between the local gap and the global magnetospheric structure.  As mentioned above,
a generic feature of magnetically driven outflows from a Kerr black hole is a plasma double-flow that emanates from a stagnation surface located between
the inner and outer Alfven surfaces.    f The location of the stagnation surface is determined from a balance between the 
gravitational, centrifugal and Lorentz forces \cite{Tak90}.  In the limit of low inertia considered in this paper (where a gap forms) it depends 
very weakly on the details of plasma injection \cite{GL14}.
In general, it has a non-spherical shape \cite{GL14,BT15}, and its distance from the BH ranges from 
$r\sim 4.5 r_g$ in the equator to $r\sim 10 r_g$ along the axis, so that it is located well outside the null surface (see Figure \ref{fig:global} for illustration).  
Now, if the outer gap boundary extends beyond the stagnation surface, then accelerated particles leaving the outer gap boundary move outwards and particles
the escape through the inner gap boundary move inwards, in accord with the global plasma flow requirements.  On the other hand, if the outer
gap boundary lies below the stagnation surface, then the direction of the particle beam that escapes through the outer gap boundary is opposite to that of the plasma 
flow in the force-free section below the stagnation surface, as illustrated schematically in Figure \ref{fig:global}.  This inconsistency most likely means 
that the plasma production process must be dynamic.   As seen in Figure \ref{fig:profiles}, in all steady solutions the outer gap boundary does not extend beyond $3r_g$,
so that it is located below the stagnation surface.   This suggests that the local steady solutions derived here may be inapplicable to a global magnetosphere. 

\section{\label{sec:conc}Conclusions}
The main conclusion of this paper is that under realistic conditions, charge-starved regions in the magnetosphere of a Kerr black hole 
are expected to be inherently intermittent.   The main reasons are that (i) for realistic values of the magnetospheric current the 
pair multiplicity cannot accommodate the closure condition required by a steady gap, unless the luminosity of the external radiation source is
extremely small, and (ii) the steady gap solutions are inconsistent with the global magnetospheric structure. 
The latter reason seems to imply that in black hole outflows the entire region below the stagnation surface should be dynamic.
It is unclear at present how the intermittency of the plasma production process will affect the resultant emission.   In local gap
models the plasma production process can be sporadic, giving rise to electric current oscillations around the mean value imposed by the global 
magnetosphere with an amplitude that depends on the pair production rate.  In this case, a reduction in the amplitude of the gap oscillations is expected when the intensity of the ambient radiation 
field, that provides the dominant pair production opacity, is increased.  In global, self-consistent gap models it seems that plasma production 
should occur in cycles of pair creation bursts.    What is the fraction of the black hole spin down power that can be released in the form of high-energy radiation
in this dynamic state and how this should affect the emitted spectrum is yet an open question.    

 Intermittency is expected also in pulsar gaps under certain conditions \cite{Lev05,Tim13}, however,
the reason for this is different than in the case of rotating black holes discussed here.

\begin{acknowledgments}
This research was supported by  a grant from the Israel Science Foundation no. 1277/13.
\end{acknowledgments}

\begin{widetext}

\appendix
\section{Derivation of the generalized Gauss' law}
From the inhomogeneous Maxwell's equations,
\begin{equation}
\frac{1}{\sqrt{-g}}\partial_\mu (\sqrt{-g}F^{\nu\mu})=4\pi j^\nu,
\end{equation}
and the relation 
\begin{eqnarray}
F^{t\mu} =  g^{\mu\nu}(g^{tt}F_{t\,\nu }+g^{t\varphi}F_{\varphi\,\nu})  =\frac{1}{\alpha^2}g^{\mu\nu}(F_{\nu\, t}+\omega F_{\nu\,\varphi})
\end{eqnarray}
one obtains the generalized Gauss' law:
\begin{equation}
\frac{1}{\sqrt{-g}}\partial_\mu\left[\frac{\sqrt{-g}\, g^{\mu\nu}}{\alpha^2} (F_{\nu\, t}+\omega F_{\nu\,\varphi})      \right]=4\pi\, j^t.
\label{Gauss_app_gen}
\end{equation}
In terms of the electric field measured in a frame rotating with the flux tube, 
\begin{eqnarray}
F^\prime_{\alpha t}= F_{\alpha t}+\Omega F_{\alpha\varphi},
\end{eqnarray}
the latter equation can be written as
\begin{equation}
\frac{1}{\sqrt{-g}}\partial_\mu\left[\frac{\sqrt{-g}\, g^{\mu\nu}}{\alpha^2}F^\prime_{\nu t}\right]=4\pi(j^t-\rho_{GJ}),
\label{Gauss_app_rot}
\end{equation}
where 
\begin{equation}
\rho_{GJ}=\frac{1}{4\pi \sqrt{-g}}\partial_\mu\left[\frac{\sqrt{-g}\, g^{\mu\nu}}{\alpha^2}(\omega-\Omega)F_{\nu\varphi}\right].
\label{rho_GJ_app}
\end{equation}
For the static, axisymmetric radial gap invoked in section \ref{sec:analysis} we have $\partial_t=\partial_\varphi=0$ and $F^\prime_{\theta\,t}=0$, whereby 
the expression for $\rho_{GJ}$ reduces to Equation (\ref{rho_GJ_1D_split}), 
and Equation (\ref{Gauss_app_rot}) reduces to
\begin{equation}
\frac{1}{\Sigma}\partial_r\left(\frac{A}{\Sigma}F^\prime_{r t}\right)=4\pi(j^t-\rho_{GJ}),
\label{Gauss_app_split}
\end{equation}
where the substitutions $\sqrt{-g}=\Sigma\sin\theta$ and  $\sqrt{-g}g^{rr}/\alpha^2=(A/\Sigma)\sin\theta$ have been used.
Upon defining the electric flux as $\Phi_E= A F^\prime_{r t}/\Sigma$ and transforming to the tortoise coordinate given in Equation (\ref{tortoise_cord}),
Equation (\ref{1D_Poiss}) is obtained.

\section{Derivation of the fluid equations}
\label{app:fluid}
The plasma in the gap is treated as a two-component fluid consisting of electrons and positrons, with proper densities $n_\pm$, pressures $p_\pm$, 
specific enthalpies (per particle)  $h_\pm$, and 4-velocities  $u^\mu_\pm$, where subscript $-$ ($+$) designates the electron (positron) fluid.   In the presence of pair creation the continuity 
equation becomes,
\begin{equation}
\frac{1}{\sqrt{-g}}\partial_\mu(\sqrt{-g}n_\pm u_{\pm}^{\mu}) =Q/2,
\label{app:cont}
\end{equation}
here $Q$ denotes the pair production rate per unit volume.   
The electric 4-current is given by
\begin{equation}
j^\mu=e(n_+u_+^\mu - n_-u_-^\mu),
\end{equation}
and from Eq. (\ref{app:cont}) it is readily seen that the electric current is conserved, viz., $\partial_\mu j^\mu=0$.  In a steady gap
this implies that the current is constant inside the gap, $\nabla\cdot {\bf j}=0$.

The energy-momentum equation takes the form:
\begin{equation}
\frac{1}{\sqrt{-g}}\partial_\nu(\sqrt{-g}T_{\pm}^{\mu \nu})+\Gamma^\mu_{\, \, \,  \alpha\beta}T_{\pm}^{\alpha\beta}
=\pm en_\pm F^\mu_{\, \, \, \alpha}u_{\pm}^\alpha - S_{\pm}^\mu + Q_{\pm}^\mu,
\label{app:e-p-eq}
\end{equation}
in terms of the energy-momentum tensor
\begin{equation}
T_\pm ^{\mu \nu}=h_\pm n_\pm u_\pm^\mu u_\pm^\nu + p_\pm g^{\mu \nu}.
\end{equation}
The first term on the right hand side of Eq. (\ref{app:e-p-eq}) accounts for the work done on the fluids by electromagnetic forces,
the second term ($S_\pm^\mu$)  for radiative losses, and the third term ($Q_\pm^\mu$) is associated with pair loading via annihilation of photons.
The projection of Eq. (\ref{app:e-p-eq}) on the 4-velocity $u^\nu$ yields an equation for the change in the entropy per particle $\sigma$ (in $k_B$ units) of each fluid:
\begin{equation}
n_\pm T_\pm u_\pm ^\mu\nabla_\mu \sigma_\pm=
(S_{\pm\alpha} - Q_{\pm \alpha})u_\pm^\alpha - h_\pm Q/2,
\label{app:entropy}
\end{equation}
here $T_\pm$ is the temperature of the fluids.  By employing Eqs. (\ref{app:cont}), (\ref{app:e-p-eq})-(\ref{app:entropy}), 
and the second law, $dh-dp/n=Td\sigma$, we arrive at
\begin{equation}
n_\pm h_\pm u_\pm ^\mu\nabla_\mu u_\pm ^\nu =\pm en_\pm F^\nu_{\, \, \, \alpha}u_{\pm}^\alpha +
(- S_{\pm\alpha} + Q_{\pm \alpha} - \partial_\alpha p_\pm)(g^{\alpha\nu}+u_\pm^\alpha u_\pm^\nu),
\label{app:eq-of-mot}
\end{equation}
denoting 
$u_\pm ^\mu\nabla_\mu u_\pm ^\nu=u_\pm ^\mu\partial_\mu u_\pm ^\nu + \Gamma^\nu_{\, \, \,  \alpha\beta}u_\pm^\alpha u_\pm^\beta$.
We now make the following approximations:  First, pressure forces are expected to be small compared with the electric and radiation forces, thus
we neglect the term $\partial_\alpha p_\pm$.   Second, we assume that each fluid is approximately adiabatic, $u^\nu\nabla_\nu \sigma_\pm=0$.  
This assumption is reasonable 
if the spread in momentum is much smaller than the bulk momentum.    Under the above simplifications Eq. (\ref{app:entropy})  yields 
$(- S_{\pm\alpha} + Q_{\pm \alpha} - \partial_\alpha p_\pm)(g^{\alpha\nu}+u_\pm^\alpha u_\pm^\nu)=-S_\pm^\nu+Q_\pm^\nu-Qhu_\pm^\nu/2$.  
Third, if newly created pairs are added to the fluid with an average momentum that is roughly equal to the bulk momentum (as naively
expected from energy-momentum conservation), then  $Q_\pm^\nu-Qhu_\pm^\nu/2=0$.  With these approximations the radiative source
term is orthogonal to the fluid velocity, viz., $u_\pm^\nu S_{\pm\nu}=0$.

Next, we take the radial ($\nu=r$) component of  Eq. (\ref{app:eq-of-mot}), make use of the relation 
$u_\pm ^\mu\nabla_\mu u_{\pm r}=u_\pm ^\mu\partial_\mu u_{\pm r} - \Gamma_{\alpha r \beta}u_\pm^\alpha u_\pm^\beta$ and
the fact that $u_r\Gamma^r_{\, \, \alpha\beta}= u^r\Gamma_{r\alpha\beta}$, and note that for the invoked gap geometry $u^\mu\partial_\mu=u^r\partial_r$,
to get
\begin{equation}
\partial_r(u_\pm^2/2)=\frac{1}{2}(u_r\partial_r u^r+u^r\partial_r u_r) =-\frac{1}{2}(u_\pm^{t})^2\partial_r\alpha^2\pm\frac{e}{h_\pm}F_{rt}u_\pm^t+s_{\pm r}u_\pm^t
\label{app:eq-of-mot2},
\end{equation}
where $s^r_\pm=-S^r_\pm/(u_\pm^t n_\pm h_\pm)$ and $s_{\pm r}=g_{rr}s_\pm^r$.
Noting that $\partial_\varphi$ is a Killing vector we further have 
\begin{equation}-
u^\mu \nabla_\mu u_{\pm\varphi}=\pm \frac{e}{n_\pm h_\pm}F_{\varphi \nu}u^\nu + s_{\pm\varphi}=s_{\pm\varphi},
\end{equation}
since $F_{\varphi \nu}u^\nu=F_{\varphi r}u^r=0$ for the split monopole geometry invoked in our gap model.   Neglecting the toroidal 
component of the radiative force, $s_{\pm\varphi}=0$, which is reasonable for the assumed isotropic radiation field, implies
that the angular momentum of each fluid is conserved:
$u_{\pm\varphi}=g_{\varphi\varphi}(u^\varphi-\omega u^t)={\rm const}$.  For simplicity, we take the angular momentum of the fluids
to be zero (although our analysis can be readily extended to fluids with nonzero angular momentum).    Then, $u_\pm^\varphi=\omega u^t$, and 
from the normalization condition $u_\mu u^\mu=-1$ we readily have $(\alpha u_\pm^t)=1+g_{rr} (u_\pm^r)^2=1+u^2_\pm$, which simply defines
the Lorentz factor of the fluid measured by a ZAMO, $\gamma_\pm=\alpha u_\pm^t$.
Upon substituting the relation $\gamma^2_\pm-1=u^2_\pm$ into Eq. (\ref{app:eq-of-mot2}), using the orthogonality condition $s_\mu u^\mu=s_{\pm t}u_\pm^t+s_{\pm r}u_\pm^r=0$, noting that $s_\pm^t=g^{tt}s_{\pm t}+g^{t\varphi}s_{\pm \varphi}=-s_{\pm t}/\alpha^2$, since we invoke $s_{\pm\varphi}=0$,  and transforming to the tortoise coordinate, we arrive at Eq. (\ref{eq_of_mot}).

\section{Radiation}
\label{app:radiation}
\subsection{Transport equation}
In terms of the absorption coefficient $\kappa_\nu$ and the emissivity $g_\nu=c^2 j_\nu/(h^4\nu^3)$, the transport equation for the 
photon distribution function, $f(x^\mu,p^\nu)$, takes the covariant form:
\begin{equation}
p^\alpha\partial_\alpha  f-\Gamma^\alpha_{\beta\gamma} p^\beta p^\gamma\frac{\partial f}{\partial p^\alpha}=  p^t(-\kappa_\nu f+g_\nu),
\end{equation}
where $\Gamma^\alpha_{~\beta\gamma}$ is the usual Christoffel  symbol.  With respect to a ZAMO frame 
defined by the tetrads 
$e_{\hat{t}}=\frac{1}{\alpha}(\partial_t+\omega\partial_\varphi)$, $e_{\hat{r}}=\frac{1}{\sqrt{g_{rr}}}\partial_r$, 
$e_{\hat{\theta}}=\frac{1}{\sqrt{g_{\theta\theta}}}\partial_\theta$, $e_{\hat{\varphi}}=\frac{1}{\sqrt{g_{\varphi\varphi}}}\partial_\varphi$,
the components of the photon momentum are $p_{\hat{a}}= e_{\hat{a}}^b p_b$, 
and $p^{\hat{a}}=\eta^{\hat{a}\hat{b}}p_{\hat{b}}$.  In this frame we define the direction 
vectors $n^{\hat{a}}=(1,\mu_p,\sin\theta_p\cos\varphi_p, \sin\theta_p\sin\varphi_p)$, where $\mu_p=\cos\theta_p$ \cite{MK86}.
Clearly $n_{\hat{a}}n^{\hat{a}}=0$, as required.  The photon momentum in this frame is $p^{\hat{a}}=\nu n^{\hat{a}}$, where henceforth 
we use units where $h=c=1$.  Note that the angle $\theta_p$ is measured with respect to the radial direction $\partial_r$.
We suppose that the photon distribution is axi-symmetric locally, that is $f$ is independent of $\varphi_p$.  Then, the transfer equation 
takes the form \cite{Lin66,MK86}
\begin{equation}
\left[n^{\hat{a}}\partial_{\hat{a}} -\gamma^{\hat{t}}_{\hat{a}\hat{b}}n^{\hat{a}}n^{\hat{b}}\,\nu\frac{\partial}{\partial\,\nu} 
+(n^{\hat{r}}\gamma^{\hat{t}}_{\hat{a}\hat{b}}-\gamma^{\hat{r}}_{\hat{a}\hat{b}})n^{\hat{a}}n^{\hat{b}}\frac{\partial}{\partial\,\mu_p}  \right]f =-\kappa_\nu f+g_\nu,
\end{equation}
in terms of the Ricci rotation coefficients
$\gamma^{\hat{a}}_{\hat{b}\hat{c}}= e^{\hat{a}}_\lambda e^\nu_{\hat{c}}(\partial_\nu e^\lambda_{\hat{b}}
+\Gamma^\lambda_{\nu\mu}e^\mu_{\hat{b}})$.

In applying the transport equation to the gamma ray emission in the gap we take $\nu=\epsilon_\gamma$, $\mu_p=\mu_\gamma$, $\varphi_p=\varphi_\gamma$, 
$f=I_\gamma(r,\epsilon_\gamma,\mu_\gamma)/\epsilon_\gamma^3$.  Since the beamed intensity is independent of $\varphi_\gamma$ we can average the transport
equation over this angle.   Using the relations
\begin{equation}
\gamma^{\hat{t}}_{\hat{a}\hat{b}}\,n^{\hat{a}}n^{\hat{b}} = n^{\hat{a}}\partial_{\hat{a}}\ln\alpha=\frac{\mu_\gamma}{\sqrt{g_{rr}}}\partial_r\ln\alpha
\end{equation}
and
\begin{equation}
\frac{1}{2\pi}\int \gamma^{\hat{r}}_{\hat{a}\hat{b}}\,n^{\hat{a}}n^{\hat{b}} d\varphi_\gamma=\gamma^{\hat{r}}_{\hat{t}\hat{t}}
+\frac{1}{2}(1-\mu_\gamma^2)(\gamma^{\hat{r}}_{\hat{\theta}\hat{\theta}}+\gamma^{\hat{r}}_{\hat{\varphi}\hat{\varphi}})
=\frac{1}{\sqrt{g_{rr}}}[\partial_r\ln\alpha - (1-\mu^2_\gamma) \partial_r\ln\sqrt{A}],
\end{equation}
one finally arrives at:
\begin{eqnarray}
n^{\hat{a}}\partial_{\hat{a}} I_\gamma - \frac{\mu_\gamma}{\sqrt{g_{rr}}}
(\partial_r\ln\alpha) \epsilon_\gamma^4\frac{\partial}{\partial\epsilon_\gamma}(I_\gamma/\epsilon_\gamma^3)
+\left[-\partial_r\ln\alpha+\frac{1}{2}\partial_r\ln\sqrt{A} \right]
\frac{(1-\mu_\gamma^2)}{\sqrt{g_{rr}}}\frac{\partial}{\partial\mu_\gamma}I_\nu=-\kappa_{pp} I_\gamma +j_\gamma.
\end{eqnarray}
To simplify our analysis we shall neglect the term $\partial_r\ln\alpha/\sqrt{g_{rr}}$ as it is merely important very near the horizon. 
we further apply the beaming approximation (\ref{beam_gamma}), note that $n^{\hat{a}}\partial_{\hat{a}}=\frac{1}{\sqrt{g_{rr}}}\partial_r$,
and integrate the later equation over the angle $\mu_\gamma$ to obtain:
\begin{equation}
\frac{1}{\sqrt{A}}\partial_{r}(\sqrt{A} I^\pm_\gamma)=\sqrt{g_{rr}}(\pm\kappa_{pp} I^\pm_\gamma \mp j^\pm_\gamma).
\end{equation}
Upon transforming to the tortoise coordinate we obtain Eq. (\ref{transfer_eq}).

\subsection{Inverse Compton emissivity}
We consider inverse Compton scattering of target radiation by a cold electron (positron) beam of comoving density $n_\pm$.
The intensity of the target radiation, as measured in the rest frame of the beam,  is denoted by $I^\prime_s(\epsilon^\prime_s,\mu^\prime_s,r,t)$,
with $\epsilon_s^\prime$, $\mu^\prime_s$ being the energy and direction of the target photons.
The comoving gamma-ray emissivity has the general form
\begin{equation}
j^\prime_{\gamma}(\epsilon^\prime_\gamma,\mu^\prime_\gamma,r,t)=n_\pm\int\frac{\epsilon^\prime_\gamma}{\epsilon^\prime_s}I^\prime_s(\epsilon^\prime_s,\mu^\prime_s,r,t)
\frac{d\sigma^\prime}{d\Omega^\prime_\gamma}\delta[\epsilon^\prime_\gamma-\epsilon^\prime_c(\epsilon^\prime_s)] d\epsilon^\prime_sd\Omega^\prime_s,
\label{app:emiss-comov}
\end{equation} 
here
\begin{equation}
\epsilon^\prime_c(\nu^\prime_s)=\frac{\epsilon^\prime_s}{1+\frac{h\epsilon^\prime_s}{m_ec^2}(1-\cos\psi)},
\end{equation}
$\psi$ is the angle between the incident and scattered photons, given by
$\cos\psi=\mu^\prime_\gamma\mu^\prime_s+\sin\theta^\prime_\gamma\sin\theta^\prime_s\cos(\varphi^\prime_\gamma-\varphi^\prime_s)$,
and
\begin{equation}
\frac{d\sigma^\prime}{d\Omega^\prime_\gamma}=\frac{3\sigma_T}{16\pi}\left(\frac{\epsilon^\prime_\gamma}{\epsilon^\prime_s}\right)^2
\left(\frac{\epsilon^\prime_\gamma}{\epsilon^\prime_s}+\frac{\epsilon^\prime_s}{\epsilon^\prime_\gamma}-\sin^2\psi \right)
\end{equation}
is the differential Klein-Nishina cross-section.  
In our model the target radiation field is taken to be isotropic in the ZAMO frame
with a power law spectrum, $I_s=I_0(r)(\epsilon_s/\nu_{min})^{-p}$, $\epsilon_{min} < \epsilon_s < \epsilon_{max}$.  
Since the Lorentz factor of the beams, $\gamma_\pm$, is extremely large, we safely assume that the target radiation field is 
completely beamed in the comoving frame.  Specifically, 
\begin{equation}
I^{\prime\pm}_s(\epsilon^\prime_s,\mu^\prime_s,r)
=\frac{4\gamma_\pm}{3}I_0(r)(\epsilon^\prime_s/2\gamma_\pm\epsilon_{min})^{-p}\delta(1\mp\epsilon^\prime_s);
\quad 2\gamma_\pm\epsilon_{min}<\epsilon^\prime_s<2\gamma_\pm\epsilon_{max},
\end{equation}
where superscript $+$ ($-$) refers to the positron (electron) beam.  
Performing the integral in Eq. (\ref{app:emiss-comov}) and noting that 
$|d\epsilon_c^\prime/d\epsilon_s^\prime|=(\epsilon_c^\prime/\epsilon_s^\prime)^2$ yields:
\begin{equation}
j^{\pm \prime}_{\gamma}(\epsilon^\prime_\gamma,\mu^\prime_\gamma,r)=\frac{\sigma_Tn_\pm}{2}I_0 (\epsilon^\prime_s/2\gamma_\pm \epsilon_{min})^{-p}
\frac{\epsilon_\gamma^\prime}{\epsilon^\prime_s}\left[\frac{\epsilon_\gamma^\prime}{\epsilon^\prime_s}
+\frac{\epsilon_s^\prime}{\epsilon^\prime_\gamma}-1+\mu_\gamma^{\prime 2}\right].
\end{equation}
Transforming back to the ZAMO frame, and recalling that $$\epsilon_\gamma^\prime/\epsilon^\prime_s=1-\epsilon_\gamma(1-\beta_\pm)\gamma_\pm(1+\mu_\gamma),
\quad \frac{\epsilon_s^\prime}{2\gamma_\pm\epsilon_{min}}=\frac{\epsilon_\gamma}{2\epsilon_{min}}\left[\frac{1-\beta_\pm\mu_\gamma}{1-\epsilon_\gamma
\gamma_\pm(1-\beta_\pm)(1+\mu_\gamma)}\right],$$ 
we have
\begin{equation}
j^\pm_{\gamma}(\epsilon_\gamma,\mu_\gamma,r)=\frac{j^{\pm\prime}_{\gamma}(\epsilon^\prime_\gamma,\mu^\prime_\gamma,r)}{[\gamma_\pm(1-\beta_\pm\mu_\gamma)]^2}=\frac{\sigma_Tn_\pm\gamma_\pm}{2} I_0(r) (\epsilon_\gamma/2 \epsilon_{min})^{-p}\,g(\epsilon_\gamma,\mu_\gamma,\gamma_\pm),
\end{equation} 
where
$$
g(\epsilon_\gamma,\mu_\gamma,\gamma_\pm))=\frac{1}{\gamma_\pm^2(1-\beta_\pm\mu_\gamma)^2}\left[\frac{2\gamma_\pm-\epsilon_\gamma(1+\mu_\gamma)}{2\gamma_\pm(1-\beta_\pm\mu_\gamma)}\right]^p\frac{\epsilon_\gamma^\prime}{\epsilon^\prime_s}\left[\frac{\epsilon_\gamma^\prime}{\epsilon^\prime_s}
+\frac{\epsilon_s^\prime}{\epsilon^\prime_\gamma}-1+\mu_\gamma^{\prime 2}\right].
$$
%
Noting that the minimum and maximum scattering angles for a given gamma ray energy are
\begin{eqnarray}
1-\beta_\pm\mu_{min}=min\left[(1+\beta_\pm), \frac{2\epsilon_{max}}{\epsilon_{\gamma}}-\frac{2\epsilon_{max}}{\gamma_\pm}\right],\\
1-\beta_\pm\mu_{max}=max\left[(1-\beta_\pm), \frac{2\epsilon_{min}}{\epsilon_{\gamma}}-\frac{2\epsilon_{min}}{\gamma_\pm}\right],\nonumber
\end{eqnarray}
and averaging the emissivities over angles, that is,
\begin{equation}
j^\pm_{\gamma}(\epsilon_\gamma,r)=\frac{1}{2}\int_{\mu_{min}}^{\mu_{max}}j^\pm_{\gamma}(\epsilon_\gamma,\mu_\gamma,r) d\mu_\gamma,
\end{equation}
one obtains, to leading order, the beamed emissivities in Eq. (\ref{j_IC_1}).

\section{Derivation of $<Q_{cur}>$}
\label{sec:app-transfer}
As argued below Eq. (\ref{kpp,cur}), at energies $\epsilon_\gamma<\epsilon_c$ the pair production opacity is much smaller than unity.  We can therefore 
neglect absorption in the transfer equation (\ref{transfer_eq}).   We can also neglect the IC emissivity since it is much smaller than the curvature 
emissivity at these energies.   The approximate solutions to Eq. (\ref{kpp,cur}), subject to the boundary conditions (\ref{bc-3}), then read:
\begin{eqnarray}
I_\gamma^+(r,\epsilon_\gamma)&=&-\frac{\sqrt{3}\alpha_f}{2\pi R_c}\gamma_+
 ~F(\epsilon_\gamma/\epsilon_c)\frac{1}{\sqrt{A(r)}}\int_r^{r_{out}}\frac{\sqrt{A}N^r_+}{ \Delta}dr^\prime,\label{app_Ig+}\\
I_\gamma^-(r,\epsilon_\gamma)&=&\frac{\sqrt{3}\alpha_f}{2\pi R_c}\gamma_- ~F(\epsilon_\gamma/\epsilon_c)
\frac{1}{\sqrt{A(r)}}\int_{r_{in}}^r\frac{\sqrt{A}N^r_-}{ \Delta}dr^\prime.
\label{app_Ig-}\
\end{eqnarray}
We note that $\sqrt{A(r)}$ changes by at most a factor of 3 across the gap, so we assume it s constant with an average 
value $<\sqrt{A}>=(\sqrt{A_{in}}+\sqrt{A_{out}})/2$, where $A_{in}\equiv A(r_{in})$ and likewise for $A_{out}$.
Recalling that $N^r_+ - N^r_- = N^r_0 =$ const, and that $\gamma^+=\gamma^-\equiv \gamma$ across most of the gap, and taking $<\sqrt{A}>$ 
instead of $\sqrt{A(r)}$, the sum of the two intensities yields
\begin{equation}
I_\gamma^+(r_{in},\epsilon_\gamma)+I_\gamma^-(r_{out},\epsilon_\gamma)=\frac{\sqrt{3}\alpha_f}{2\pi R_c}\gamma~F(\epsilon_\gamma/\epsilon_c)(-N^r_0)
\frac{1}{<\sqrt{A}>}\int_{r_{in}}^{r_{out}}\frac{\sqrt{A} dr^\prime}{ \Delta}
\end{equation}
With the crude approximation $I_\gamma^+(r,\epsilon_\gamma)+I_\gamma^-(r,\epsilon_\gamma)=
[I_\gamma^+(r_{in},\epsilon_\gamma)+I_\gamma^-(r_{out},\epsilon_\gamma)]/2$,  we obtain an expression for the average
pair production rate:
\begin{equation}
<Q_{cur}>=\frac{\sqrt{27}\alpha_f}{2 R_c}\gamma~\tau_0~(-N^r_0)A_p (\epsilon_c\epsilon_{min})^p
\frac{1}{<\sqrt{A}>}\int_{r_{in}}^{r_{out}}\frac{\sqrt{A} dr^\prime}{2 \Delta}~ \int_0^\infty x^{p-1}F(x)dx.
\end{equation}
By employing Eqs. (\ref{gam_sat}) and (\ref{eps_c_2}), choosing $p=2$ for illustration, and computing the last integral 
on the right hand side we arrive at Eq. (\ref{Q_cur_av}).

\end{widetext}


\begin{thebibliography}{99}
\bibitem{BZ77} R. D. Blandford and R. L. Znajek, Mon. Not. R. Astron. Soc.,  \textbf{179}, 433, (1977)
\bibitem{BK08} M. Barkov and S. Komissarov,  Mon. Not. R. Astron. Soc., \textbf{385}, L28 (2008)
\bibitem{LR11} A. Levinson and F. Rieger, Astrophys. J., \textit{730}, 123 (2011)
\bibitem[\protect\citeauthoryear{Levinson}{2000}]{le00} A. Levinson, PhRvL, \textbf{85}, 912 (2000)
\bibitem[\protect\citeauthoryear{Hirotani \& Pu}{2016}]{HP16} K.~Hirotani, Pu H.-Y., 2016, Astrophys. J., \textbf{818}, 50 (2016)
\bibitem[\protect\citeauthoryear{Hirotani et al.}{2016}]{Hir16} K.~Hirotani, H.-Y. Pu, L.~C.-C. Lin, H.-K.~Chang, M. Inoue, A.~K.~H. Kong, S.~Matsushita and P.-H.~T. Tam, Astrophys. J., \textbf{833}, 142 (2016)
\bibitem[\protect\citeauthoryear{Hirotani et al.}{2017}]{Hir17} K.~Hirotani, H.-Y.~ Pu, L.~C.-C. Lin, A.~K.~H. Kong, S.~Matsushita, K.~Asada, H.-K. Chang and P.-H.~T. Tam, Astrophys. J., \textbf{845}, 77 (2017)
\bibitem[\protect\citeauthoryear{Lin et al.}{2017}]{Lin17}  L.~C.-C. Lin, H.-Y. Pu, K.~Hirotani, A.~K.~H.~Kong, S.~Matsushita, H.-K. Chang, M.~Inoue and P.-H.~T. Tam, Astrophys. J., \textbf{845}, 40 (2017)
\bibitem[\protect\citeauthoryear{Neronov \& Aharonian}{2007}]{NA07} A.~Neronov, F.~A.~Aharonian, Astrophys. J., \textbf{671}, 85 (2007)
\bibitem[\protect\citeauthoryear{Rieger}{2011}]{Rie11} F.~M.~Rieger, Int. J. Mod. Phys. D, \textbf{20}, 1547 (2011)
\bibitem[\protect\citeauthoryear{Broderick \& Tchekhovskoy}{2015}]{BT15} A.~E.~Broderick and A. Tchekhovskoy,  Astrophys. J., \textbf{809}, 97 (2015)
%
\bibitem[\protect\citeauthoryear{Aharonian et al.}{2003}]{ahr03} F.~A.~Aharonian, et al., Astron. Astrophys., \textbf{403}, L1 (2003)
\bibitem[\protect\citeauthoryear{Acciari et al.}{2009}]{acc09} V.~A.~Acciari, et al.,  Sci, \textbf{325}, 444 (2009)
\bibitem[\protect\citeauthoryear{Aleksi{\'c} et al.}{2014}]{alk14} J.~Aleksi{\'c}, et al.,  Sci, \textbf{346}, 1080 (2014)
%
\bibitem[\protect\citeauthoryear{Chen \& Beloborodov}{2014}]{CB14}  A.~Y. Chen and A.~M.~ Beloborodov,  Astrophys. J., \textbf{795}, L22 (2014)
\bibitem[\protect\citeauthoryear{Cerutti et al.}{2015}]{CPPS15} B.~Cerutti, A.~Philippov, K.~Parfrey and A.~Spitkovsky, Mon. Not. R. Astron. Soc., \textbf{448}, 606 (2015)
\bibitem[\protect\citeauthoryear{Philippov, Spitkovsky, \& Cerutti}{2015}]{PSC15} A.~A.~Philippov, A.~Spitkovsky and B.~ Cerutti, Astrophys. J., \textbf{801}, L19 (2015)
%
\bibitem[\protect\citeauthoryear{Globus \& Levinson}{2014}]{GL14} N.~Globus and A.~ Levinson,  Astrophys. J., \textbf{796}, 26 (2014)
\bibitem[\protect\citeauthoryear{Globus \& Levinson}{2013}]{GL13} N.~Globus and A.~ Levinson,  PhRvD, \textbf{88}, 084046 (2013)
\bibitem[\protect\citeauthoryear{Narayan \& Yi}{1995}]{NY95} R.~Narayan and I.~ Yi,  Astrophys. J., \textbf{452}, 710 (1995)
\bibitem[\protect\citeauthoryear{Blandford \& Levinson}{1995}]{BL95} R.~D.~ Blandford and  A.~ Levinson, Astrophys. J., \textbf{441}, 79 (1995)
\bibitem[\protect\citeauthoryear{Rybicki \& Lightman}{1979}]{RL79} G.~B.~ Rybicki and A.~P.~ Lightman, Radiative Processes in Astrophysics
(New York: Wiley) (1979)
\bibitem[\protect\citeauthoryear{Gould \& Schr{\'e}der}{1967}]{GS67} R.~J.~ Gould and G.~P.~ Schr{\'e}der, PhRv, \textbf{155}, 1404 (1967)
\bibitem[\protect\citeauthoryear{Morita \& Kaneko}{1986}]{MK86} K.~Morita and N.~ Kaneko,  Astrophys. \& Space Sci., 121, \textbf{105} (1986)
\bibitem[\protect\citeauthoryear{Lindquist}{1966}]{Lin66} R.~W.~ Lindquist, AnPhy, \textbf{37}, 487 (1966)
\bibitem[\protect\citeauthoryear{Takahashi}{1990}]{Tak90} M. Takahashi, S. Nitta, Y. Tatematsu \& A. Tomimatsu, Astrophys. J., \textbf{363}, 206 (1990)
%
\bibitem[\protect\citeauthoryear{Levinsoni}{2005}]{Lev05} A. Levinson, D. Melrose, A. Judge \& L. Qinghuan, Astrophys. J., \textbf{631}, 456 (2005)
\bibitem[\protect\citeauthoryear{Timokhini}{2013}]{Tim13}  A.~N. Timokhin \& J. Arons,  Mon. Not. R. Astron. Soc., \textbf{429}, 20 (2013)
\end{thebibliography}
%


\end{document}